\PassOptionsToPackage{dvipsnames}{xcolor}
\documentclass[10pt,twocolumn,letterpaper]{article}

\usepackage[pagenumbers]{iccv_arxiv} 

\usepackage{multirow, pifont}
\usepackage[pagebackref,breaklinks,colorlinks,allcolors=iccvblue]{hyperref}
\usepackage[accsupp]{axessibility} 
\usepackage[table]{xcolor}
\usepackage{pdfpages}
\definecolor{iccvblue}{rgb}{0.21,0.49,0.74}

\def\figurePath{fig/}

\def\myfigureW#1#2#3#4{\begin{figure}[tb]\centering\includegraphics[width = #4\linewidth]{\figurePath#2}\caption{#3}\label{fig:#1}\end{figure}}
\def\mycfigure#1#2#3{\begin{figure*}[tb]\centering\includegraphics*[clip, width = \linewidth]{\figurePath#2}\caption{#3}\label{fig:#1}\end{figure*}}
\def\mycfigureW#1#2#3#4{\begin{figure*}[tb]\centering\includegraphics*[clip, width = #4\linewidth]{\figurePath#2}\vspace{-2mm}\caption{#3}\vspace{-2mm}\label{fig:#1}\end{figure*}}

\newcommand{\revise}[1]{\textcolor{black}{#1}}

\def\eg{\emph{e.g.,~}}

\newcommand{\figref}[1]{Fig.~\ref{#1}}%
\newcommand{\tabref}[1]{Tab.~\ref{#1}}%
\newcommand{\secref}[1]{Sec.~\ref{#1}}
\renewcommand{\eqref}[1]{Eqn.~(\ref{#1})}

\definecolor{myyellow}{rgb}{1, 1, 0.7}
\definecolor{myorange}{rgb}{1, 0.85, 0.7}
\definecolor{myred}{rgb}{1, 0.7, 0.7}
\newcommand{\reducedstrut}{\vrule width 0pt height 1.05\ht\strutbox depth 1.0\dp\strutbox\relax}
\newcommand{\sota}[1]{%
  \begingroup
  \setlength{\fboxsep}{0pt}%
  \colorbox{myred}{\reducedstrut#1\/}%
  \endgroup
}
\newcommand{\subsota}[1]{%
  \begingroup
  \setlength{\fboxsep}{0pt}%
  \colorbox{myorange}{\reducedstrut#1\/}%
  \endgroup
}

\def\paperID{7520} 
\def\confName{ICCV}
\def\confYear{2025}

\newcommand{\customfootnotetext}[2]{{
  \renewcommand{\thefootnote}{#1}
  \footnotetext[0]{#2}}}

\begin{document}

\title{Gaussian Splatting with Discretized SDF for Relightable Assets}

\author{
  Zuo-Liang Zhu\textsuperscript{1} \quad
  Jian Yang\textsuperscript{1$\dagger$} \quad
  Beibei Wang\textsuperscript{2$\dagger$} \\
  \textsuperscript{1}Nankai University \quad \textsuperscript{2}Nanjing University 
}
\twocolumn[{%
\renewcommand\twocolumn[1][]{#1}%
\maketitle
\begin{center}
    \centering
    \captionsetup{type=figure}
    \includegraphics[width=0.95\linewidth]{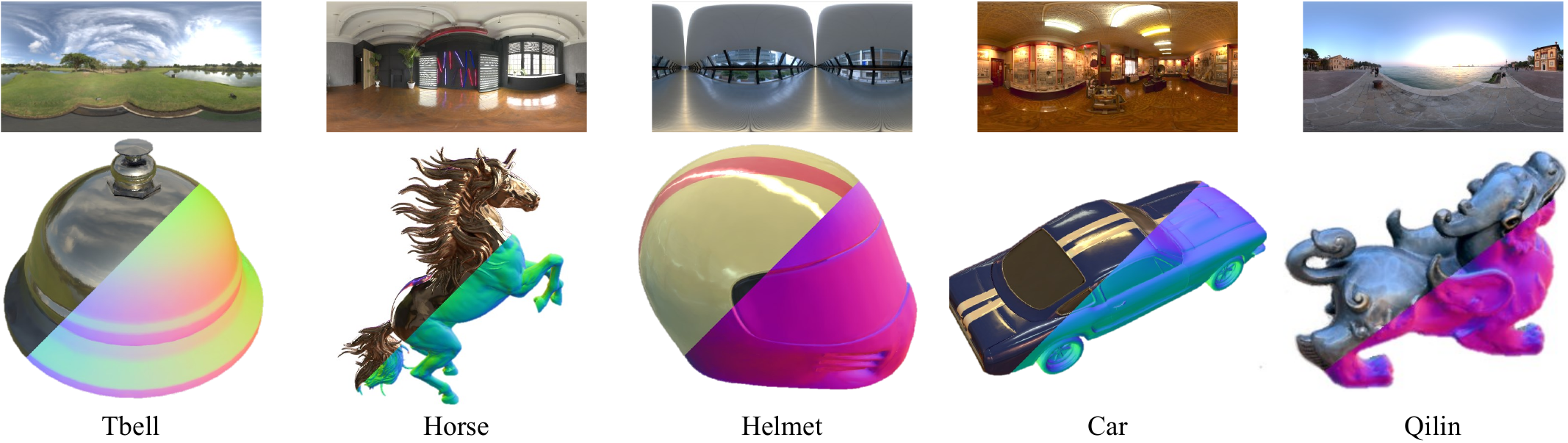}
    \caption{We present a relightable Gaussian splatting framework that introduces a discretized SDF to promote decomposition quality. We show five relighting results (left) and their normal estimation (right), including `Tbell' and `Horse' from NeRO~\cite{liu_2023_nero}, `Helmet' and `Car' from Ref-NeRF~\cite{verbin_2022_refnerf}, and `Qilin' from NeILF++~\cite{zhang_2023_neilfpp}, where `Qilin' is a real scene. Our method demonstrates a robust decomposition of geometry and material for various objects, leading to photo-realistic object relighting.}
    \label{fig:teaser}
\end{center}
}]
\customfootnotetext{$\dagger$}{Corresponding author.}
\customfootnotetext{1}{PCA Lab, VCIP, College of Computer Science, Nankai University}
\customfootnotetext{2}{School of Intelligence Science and Technology, Nanjing University, Suzhou, China.}
\begin{abstract}

3D Gaussian splatting (3DGS) has shown its detailed expressive ability and highly efficient rendering speed in the novel view synthesis (NVS) task. The application to inverse rendering still faces several challenges, \revise{as the discrete nature of Gaussian primitives makes it difficult to apply geometry constraints.} Recent works introduce the signed distance field (SDF) as an extra continuous representation to regularize the geometry defined by Gaussian primitives. It improves the decomposition quality, at the cost of increasing memory usage and complicating training. 
Unlike these works, we introduce a \textbf{discretized SDF} to represent the continuous SDF in a discrete manner by encoding it within each Gaussian using a sampled value. This approach allows us to link the SDF with the Gaussian opacity through an SDF-to-opacity transformation, enabling rendering the SDF via splatting and avoiding the computational cost of ray marching.
The key challenge is to regularize the discrete samples to be consistent with the underlying SDF, as the discrete representation can hardly apply the gradient-based constraints (\eg Eikonal loss). For this, we project Gaussians onto the zero-level set of SDF and enforce alignment with the surface from splatting, namely a projection-based consistency loss. 
Thanks to the discretized SDF, our method achieves higher relighting quality, while requiring no extra memory beyond GS and avoiding complex manually designed optimization. The experiments reveal that our method outperforms existing Gaussian-based inverse rendering methods. Our code is available at \url{https://github.com/NK-CS-ZZL/DiscretizedSDF}.

\end{abstract}    
\section{Introduction}
\label{sec:intro}

While creating relightable 3D assets from multi-view images is valuable for many downstream tasks (\eg augmented reality, virtual reality), it is challenging due to the ambiguity from decomposing geometry, material, and light.

A key point for decomposing these factors is the regularization of geometry, as high-quality geometry is the prerequisite to estimate reasonable material and light. The signed distance field (SDF) has been introduced into neural radiance field (NeRF)~\cite{mildenhall_2020_nerf} as an effective geometric prior, as NeuS~\cite{wang_2021_neus}. Due to SDF constraints, several methods~\cite{li_2024_tensosdf, zhang_2022_mii, sun_2024_neuralpbir} have shown a robust decomposition of geometry and material. Unfortunately, these methods require long training and rendering time, due to the expensive ray-marching operation. More recently, 3D Gaussian splatting (3DGS)~\cite{kerbl_2023_3dgs} achieves more detailed appearance modeling while enabling real-time rendering and fast training. It has also been introduced into inverse rendering~\cite{zhu_2024_gsror, gao_2023_relightablegs, liang_2024_gsir, jiang_2024_gaussianshader}. However, the decomposition in the GS framework becomes more problematic, due to insufficient geometric constraints. Several works~\cite{yu_2024_gsdf, zhu_2024_gsror, zhang_2024_gspull, wu_2024_deferredgs} optimize an extra SDF network with Gaussians jointly for geometry regularization. While these designs improve the geometry quality, they increase the memory usage and require complex optimization strategies (\eg warm-up~\cite{yu_2024_gsdf}, or multi-stage training~\cite{zhu_2024_gsror}) to balance two representations.

\revise{In this paper, we introduce a \emph{discretized SDF} to represent the continuous SDF in a discrete manner by encoding it within each Gaussian using a sampled value. This design enables effective geometry regularization in Gaussian-based relighting frameworks while maintaining simplicity and memory efficiency. We link the SDF with the Gaussian opacity through an SDF-to-opacity transformation, allowing rendering the SDF via splatting and avoiding the computational cost of ray marching.} The key problem is to regularize \revise{SDF with the discrete samples}, as the gradient-based losses used to regularize SDF (\eg Eikonal loss~\cite{gropp_2020_eik}) are unfeasible for the discrete samples. For this, we project Gaussians onto the zero-level set of SDF and enforce alignment with the surface from splatting, namely projection-based consistency loss, proven to be an approximation of the Eikonal loss. This way, we unify the representation between SDF and Gaussian, benefiting from the flexibility of the Gaussian primitive and the robustness of SDF while maintaining simplicity, without introducing extra representation.
 \revise{Besides, inspired by previous SDF works~\cite{wang_2021_neus, yariv_2021_volsdf}, we propose a spherical initialization tailored for Gaussian splatting}, which avoids falling into local minima and improves the robustness of the model.

As a result, our method provides high-quality and robust decomposition for objects across various materials, enables realistic relighting with only 20\% of memory usage compared to previous methods~\cite{zhu_2024_gsror}, and simplifies the optimization process, thanks to the unified representation.
In summary, our main contributions include 
\begin{itemize} 
    \item a novel regularization in the GS framework with an \revise{discretized SDF} to improve decomposition quality,
    \item a projection-based loss to regularize discretized SDF, proven as an approximation of the Eikonal loss, and
    \item \revise{extensive experiments show our method outperforms existing Gaussian-based methods with diverse metrics.}
\end{itemize}

\section{Related works}
\label{sec:rel_work}

\subsection{Inverse rendering}

Inverse rendering aims to decompose geometry, material, and light from multi-view RGB images. One group of methods adopts the geometry formulation from NeRF~\cite{mildenhall_2020_nerf} in a density~\cite{jin_2023_tensoir, srinivasa_2021_nerv, boss_2021_nerd, boss_2021_neuralpil} or SDF manner~\cite{zhang_iron_2022, zhang_2021_physg, bi_2020_neural, sun_2024_neuralpbir, zhang_2022_mii, wang_2024_inverse}. Due to the well-constrained geometry of the SDF, SDF-based methods can decompose the material and geometry robustly. Especially, NeRO~\cite{liu_2023_nero} can handle challenging reflective objects by explicitly integrating the rendering equation. TensoSDF~\cite{li_2024_tensosdf} 
uses roughness-aware integration of radiance and reflectance fields to manage diverse materials. However, these methods rely on expensive ray marching and suffer from slow training speed with high memory usage in training. Unlike prior works, our discretized SDF ensures robust decomposition across various materials, while maintaining a fast training speed and low memory costs.

More recently, Gaussian-based inverse rendering methods have emerged. These methods regularize geometry from Gaussians variously. GShader~\cite{jiang_2024_gaussianshader} and GIR~\cite{shi_2023_gir} link the shortest axis of Gaussian with the normal. GShader, R3DG~\cite{gao_2023_relightablegs}, and GS-IR~\cite{liang_2024_gsir} introduce a loss to align the Gaussian normal with the depth-derived one. Later, SDF is introduced into this field to regularize geometry. DeferredGS~\cite{wu_2024_deferredgs} distills the normal from the pretrained SDF to refine the normal from Gaussians. GS-ROR~\cite{zhu_2024_gsror} utilizes the normal and depth as a bridge to optimize the Gaussian and SDF jointly. However, introducing an extra SDF network makes the model larger and harder to optimize. 

Instead of using additional continuous representation for SDF, our work aims to directly encode discrete values of SDF samples on Gaussian primitives and regularize the geometry via this discretized SDF. In this way, our framework benefits from both GS and SDF within a unified representation, leading to a more lightweight and efficient model.

\mycfigure{overview}{overview_iccv.pdf}{Overview of our framework. \revise{Our model discretizes the underlying continuous SDF into discrete samples and encodes the SDF values of samples on Gaussian primitives.} We propose the median loss to facilitate convergence and projection-based consistency loss to regularize the discrete values of SDF samples. Besides, we design a spherical initialization for foreground objects to avoid local minima.}

\subsection{Learning SDF from multi-view images}

SDF is a fundamental surface representation~\cite{mayost_2014_sdfapp} adopted by various tasks (e.g., surface reconstruction, inverse rendering). Particularly, NeuS~\cite{wang_2021_neus} that links the SDF with the density in NeRF~\cite{mildenhall_2020_nerf} became mainstream for learning SDF from multi-view images. In their framework, the SDF is stored by multi-layer perceptrons~\cite{yariv_2021_volsdf, zhang_2022_hes} or grid-like representations~\cite{li_2024_tensosdf, rosu_2023_permutosdf, wang2023neus2, li_2023_neuralangelo}, which is continuous and convenient to obtain the gradient to regularize the SDF~\cite{gropp_2020_eik, zhang_2022_hes, wang_2023_neuralsingular}.

Recently, 3DGS~\cite{kerbl_2023_3dgs} became a new framework for the multi-view stereo, providing impressive real-time novel view synthesis (NVS) results. However, due to its discrete nature, reconstructing an accurate and continuous surface is challenging. To address this, some works optimize the SDF with 3DGS jointly. NeuSG~\cite{chen_2023_neusg} align the SDF gradient with the Gaussian normal. 3DGSR~\cite{yu_2024_3dgsr} builds an SDF-to-opacity transformation and regularizes the depth along with normal from Gaussians by the ones from SDF. GSDF~\cite{yu_2024_gsdf} utilizes the depth from Gaussians to accelerate the sampling and manipulate densifying/pruning via SDF. GS-Pull~\cite{zhang_2024_gspull} introduces the pulling operation to align the zero-level set and the Gaussian position for an accurate surface. 
All previous works utilize an extra representation while \revise{our work employs a single representation that uses Gaussian primitives as discrete samples from SDF to regularize geometry.}

\section{Method}

In this section, we briefly review Gaussian splatting (\secref{ssec:prelim}) and propose a discretized SDF within the Gaussian splatting framework for inverse rendering (\secref{ssec:sdf_gs}). Then, we present a projection-based loss to constrain the geometry when common gradient-based regularization becomes unfeasible for the discrete representation (\secref{ssec:proj}). Finally, we propose a spherical initialization to avoid the geometry trapping into local minima (\secref{ssec:init}).

\subsection{Preliminaries}
\label{ssec:prelim}

\noindent
\textbf{2DGS.} 2DGS~\cite{huang_2024_2dgs} flattens the Gaussian primitive into a 2D disk defined by two tangent vectors ($t_u, t_v$) and their corresponding scaling ($s_u, s_v$), whose normal is $n=t_u\times t_v$. They compute the \revise{ray-Gaussian response} $\alpha_i$ by a ray-splat intersection and follow the rendering pipeline in 3DGS as:
\begin{equation}
    C = \sum^{n}_{i=0}{c_i\alpha_i\prod_{j=1}^{i-1}(1-\alpha_j)},
    \label{eq:gsblend}
\end{equation}
where $c_i, \alpha_i$ is the color and the ray-Gaussian response.

\noindent
\textbf{Relightable Gaussian splatting.} To enable relighting, the color of Gaussian is replaced by physically based rendering (PBR) attributes (\eg albedo, roughness). Two existing pipelines differ in the order of blending and evaluating the rendering equation, namely forward and deferred shading. The forward one renders the color for each Gaussian and blends them, while the deferred one blends the PBR attributes first and renders each pixel.

\subsection{Relightable GS with a discretized SDF}
\label{ssec:sdf_gs}

\noindent
Previous works~\cite{zhang_2024_gspull,ye_2024_3dgsdr,yu_2024_gsdf} introduced an extra SDF representation (e.g., a network) into the GS framework as geometry constraints to improve decomposition quality, leading to higher memory costs and more complex optimization strategies. Hence, the main question is how to unify the Gaussian primitive and the SDF to avoid these drawbacks. To this end, we propose the discretized SDF into the Gaussian splatting as a geometry prior, as shown in \figref{fig:overview}.

We first describe our baseline solution and introduce our discretized SDF. Our framework is built on 2DGS~\cite{huang_2024_2dgs} with the deferred shading, as the 2D Gaussian is robust for geometry~\cite{huang_2024_2dgs} and deferred shading is more stable across diverse materials~\cite{wu_2024_deferredgs, zhu_2024_gsror}. Besides, we choose the Principled Bidirectional Reflectance Distribution Function (BRDF)~\cite{burley_2012_physically} with split-sum approximation~\cite{karis_2013_splitsum} as our shading model.

Next, we propose \revise{the discretized SDF} into our framework. Instead of using an extra representation for SDF, we encode an SDF value on the Gaussian as an extra attribute, besides PBR attributes, scaling, and quaternion. Then, the opacity in each Gaussian is derived from the SDF value via an SDF-to-opacity transformation:
\begin{equation}
    o_i =\mathcal T_{\gamma}(s_i)= \frac{4\cdot e^{-\gamma s_i}}{(1+e^{-\gamma s_i})^2},
\end{equation}
where $o_i$ is the opacity, $s_i$ is the SDF value of $i$-th Gaussian, and $\gamma$ is a global learnable parameter to control the variance of the transformation. Note that the opacity of each Gaussian is computed from the SDF, rather than directly learned. 
With the transformation, Gaussians can be splatted the same as in previous works~\cite{yu_2024_3dgsr, huang_2024_2dgs}, and the Gaussian attributes together with the parameter of the transformation can be optimized. However, we observe a slow convergence, as the joint optimization of the transformation parameter $\gamma$ and the SDF value $s_i$ spans a large search space.
\myfigureW{s2o}{s2o_iccv}{The SDF-to-opacity transformation $\mathcal T_{\gamma}$ for splatting is a bell-shaped function ranging from zero to one. The proposed median loss $\mathcal L_{\gamma}$ drags $\mathcal T_{\gamma}$ (red line) to narrower $\mathcal T_{\gamma_{\rm m}}$ (blue line).}{0.95}
\myfigureW{contvsdis}{contvsdis_iccv}{The difference between continuous and discretized SDF in regularization.
 For discrete samples, we can only obtain the gradient direction of SDF,\protect\footnotemark  so the Eikonal loss is infeasible. Besides, since discretized SDF does not support the query of SDF value at the projected point, we cannot constrain projected points on the surface directly and thus use depth as a bridge to achieve that.}{0.95}
\footnotetext{An ideal SDF has a gradient magnitude of one, which does not hold during optimization, and discrete-sample magnitudes cannot be directly measured. Thus, only the direction on Gaussian is accessible.}

\revise{To mitigate the challenges in the joint learning of $\gamma$ and $s_i$, we found the median of unsigned distance $|s|_{\rm m}$ across all Gaussian primitives is an effective indicator for convergence. Specifically, a high opacity $o_{\rm m}=\mathcal T_{\gamma_{\rm m}}(|s|_{\rm m})$ at the median distance indicates the transformation should be narrowed, as shown in \figref{fig:s2o}. Hence, we link the transformation parameter to the median $|s|_{\rm m}$ and empirically encourage the opacity $o_{\rm m}$ below $0.5$. Accordingly, we derive $\gamma_{\rm m}$ from $\mathcal T_{\gamma_{\rm m}}(|s|_{\rm m})=0.5$ to meet this condition:}

\begin{equation}
    \gamma_{\rm m} = -\frac{\log(3-2\sqrt{2})}{|s|_{\rm m}},
\end{equation}
which is used to guide the transformation parameter $\gamma$ with our proposed \emph{SDF-to-opacity transformation median loss}:
\begin{equation}
    \mathcal L_{\gamma} =\max(\gamma_{\rm m}-\gamma, 0). 
\end{equation}
Note that, $\gamma_m$ is used as a guide of $\gamma$ to quickly narrow the transformation. After it reaches $\mathcal T_{\gamma_{\rm m}}$ defined by $\gamma_{\rm m}$, we do not force the transformation $\mathcal T_{\gamma}$ identical to $\mathcal T_{\gamma_{\rm m}}$ and allow it to be sharper. In this way, the Gaussians can quickly approach the surface and converge into a sharp distribution.

\myfigureW{proj}{proj_iccv}{The projection-based consistency loss minimizes the difference between the zero-level set of SDF and the alpha-blended surface defined by Gaussians. The exceptional Gaussians whose difference is larger than $\varepsilon$ are discarded.}{0.95}

\subsection{Projection-based consistency loss}
\label{ssec:proj}

A signed distance function $f$ meets the Eikonal condition $|\nabla f| = 1$, indicating the gradient magnitude to position should be constantly one, which is usually used to constrain an implicit SDF, as the gradient $\nabla f$ can be computed easily from a continuous SDF. However, it is unclear how to obtain it for SDF samples encoded on Gaussians. As shown in \figref{fig:contvsdis}, only the gradient direction on Gaussian is available, which is the normal direction $n=\frac{\nabla f}{|\nabla f|}=t_u\times t_v$. \revise{Inspired by Neural-Pull~\cite{ma_2021_np}}, we design a new constraint for the SDF samples tailored with Gaussians by projecting Gaussians to the surface defined by the zero-level set, enforcing the consistency between the projected point and alpha-blended surface defined by Gaussians, namely \emph{projection-based consistency loss}, proven to be an approximation of the Eikonal condition. \revise{Note that although our design shares a similar theory with Neural-Pull, we do not use an extra network for SDF and have no ground-truth point cloud as supervision.}
Specifically, we project a Gaussian $g_i$ onto the zero-level set, leading to a projected point $\mu_{\rm proj}$:
\begin{equation}
    \mu_{\rm proj} = \mu_i - s_i \frac{\nabla f_{i}}{|\nabla f_{i}|},
\end{equation}
where $\mu_i$, $s_i$, $\nabla f_{i}$ are the position, SDF value and SDF gradient of $g_i$, respectively. The projected point should align with the alpha-blended surface from Gaussians. For this, we ensure this property by minimizing their difference. Here, we use depth rather than position, due to simplicity, leading to the following difference $\epsilon_{\rm proj} = |D_{\rm agg} - D_{\rm proj}|$ where $D_{\rm agg}$ is the alpha-blended or aggregated depth, and $D_{\rm proj}$ is the depth of projected point $\mu_{\rm proj}$ under the same view.
\begin{table*}[t]
    \centering
\resizebox{\linewidth}{!}{
    \renewcommand{\arraystretch}{1.2}
    \begin{tabular}{lccccc}
    \hline
            & GShader & GS-IR & R3DG & GS-ROR & Ours  \\
            & PSNR / SSIM / LPIPS & PSNR / SSIM / LPIPS & PSNR / SSIM / LPIPS & PSNR / SSIM / LPIPS & PSNR / SSIM / LPIPS  \\
       \hline
       Angel  & 17.49 / 0.8336 / 0.1901 & 15.64 / 0.6126 / 0.1428 & 16.65 / 0.8013 / 0.1181 & \subsota{20.81} / \subsota{0.8775} / \subsota{0.0858} & \sota{22.03} / \sota{0.8919} / \sota{0.0819} \\
       Bell   & 19.01 / 0.8804 / 0.5203 & 12.61 / 0.2807 / 0.1431 & 16.15 / 0.8391 / 0.1329 & \subsota{24.49} /  \subsota{0.9267} / \sota{0.0795} & \sota{24.67} / \sota{0.9280} / \subsota{0.0842} \\
       Cat    & 16.00 / 0.8642 / 0.1591 & 18.04 / 0.7907 / 0.1171 & 17.49 / 0.8503 / 0.1114 & \subsota{26.28} / \subsota{0.9421} / \sota{0.0596} & \sota{26.48} / \sota{0.9374} / \subsota{0.0661} \\ 
       Horse  & 22.49 / 0.9262 / 0.1606 & 17.40 / 0.7270 / 0.0866 & 20.63 / 0.8832 / 0.0498 & \subsota{23.31} / \subsota{0.9376} / \subsota{0.0356} & \sota{24.01} / \sota{0.9481} / \sota{0.0351} \\
       Luyu   & 15.62 / 0.8254 / 0.1391 & 19.00 / 0.7727 / 0.1094 & 17.47 / 0.8168 / 0.1057 & \subsota{22.61} / \subsota{0.8995} / \sota{0.0672} & \sota{23.80} / \sota{0.9017} / \subsota{0.0699} \\ 
       Potion & 12.33 / 0.7575 / 0.2142 & 16.37 / 0.7051 / 0.1614 & 14.99 / 0.7799 / 0.1832 & \subsota{25.67} / \subsota{0.9175} / \sota{0.0937} & \sota{27.31} / \sota{0.9280} / \subsota{0.0982} \\
       Tbell  & 14.42 / 0.8007 / 0.3368 & 14.35 / 0.5419 / 0.1970 & 15.99 / 0.7965 / 0.1877 & \subsota{22.80} / \subsota{0.9180} / \sota{0.0953} & \sota{23.66} / \sota{0.9191} / \subsota{0.0981} \\
       Teapot & 18.21 / 0.8560 / 0.1716 & 16.63 / 0.7646 / 0.1312 & 17.36 / 0.8389 / 0.1194 & \subsota{21.17} / \subsota{0.8932} / \subsota{0.0982} & \sota{24.19} / \sota{0.9293} / \sota{0.0760} \\
       \hline
       Mean   & 16.95 / 0.8430 / 0.2365 & 16.26 / 0.6494 / 0.1361 & 17.09 / 0.8258 / 0.1260 & \subsota{23.39} / \subsota{0.9140} / \subsota{0.0769} & \sota{24.52} / \sota{0.9229} / \sota{0.0762} \\
       \hline
       Training  & 0.5h & 0.5h & 1h  &    1.5h   &   1h  \\
       Ren. FPS   & 50 & 214  & 50   &    208   &  143  \\
       Memory  &    4G    &   8G   &  20G & 22G & 4G \\
       \hline
        
    \end{tabular}
}    \vspace{-2mm}
    \caption{The quantitative comparison with Gaussian-based methods on the Glossy Blender dataset in terms of PSNR$\uparrow$, SSIM$\uparrow$, and LPIPS$\downarrow$. Numbers in \sota{red} indicate the best performance, and numbers in \subsota{orange} indicate the second best.}
\label{tab:glossy_relit}
\end{table*}

\begin{table*}[t]
    \centering
\resizebox{\linewidth}{!}{
    \renewcommand{\arraystretch}{1.2}
    \begin{tabular}{lccccc}
    \hline
            & GShader & GS-IR & R3DG & GS-ROR & Ours  \\
            & PSNR / SSIM / LPIPS & PSNR / SSIM / LPIPS & PSNR / SSIM / LPIPS & PSNR / SSIM / LPIPS & PSNR / SSIM / LPIPS  \\
       \hline
        Armad.    & 22.86 / 0.9280 / 0.0821 & 28.31 / 0.9121 / 0.0591 & 30.76 / 0.9526 / 0.0649 &  \sota{31.33} / \subsota{0.9593} /  \sota{0.0482} &  \subsota{31.05} / \sota{0.9621} / \subsota{0.0536} \\
       Ficus     & 24.61 / 0.9491 / 0.0956 & 24.39 / 0.8762 / 0.0360 & \subsota{27.23} / \subsota{0.9637} / \subsota{0.0437} & 26.28 / 0.9542 / 0.0439 &  \sota{27.85} / \sota{0.9639} / \sota{0.0390} \\
       Hotdog    & 17.45 / 0.8838 / 0.1408 & 22.36 / 0.8931 / 0.0875 & 24.59 / 0.9162 / 0.1275 & \subsota{25.21} / \subsota{0.9307} / \subsota{0.0771} & \sota{26.23} / \sota{0.9360} / \sota{0.0746} \\ 
       Lego      & 13.41 / 0.7904 / 0.1193 & 23.79 / 0.8750 / 0.0879 & 22.49 / 0.8682 / 0.1423 &  \subsota{25.46} / \subsota{0.9083} / \sota{0.0724} & \sota{25.81} / \sota{0.9087} /  \subsota{0.0791} \\
       \hline
       Mean      & 19.58 / 0.8878 / 0.1095 & 24.71 / 0.8891 / 0.0676 & 26.27 / 0.9252 / 0.0946 & \subsota{27.07} / \subsota{0.9381} / \sota{0.0604} & \sota{27.78} / \sota{0.9427} / \subsota{0.0626} \\
       \hline
        
    \end{tabular}
}    
\vspace{-2mm}
\caption{The quantitative comparison with Gaussian-based methods on the TensoIR Synthetic dataset in terms of PSNR$\uparrow$, SSIM$\uparrow$, and LPIPS$\downarrow$. Numbers in \sota{red} indicate the best performance and numbers in \subsota{orange} indicate the second best.}
\vspace{-2mm}
\label{tab:tir_relit}
\end{table*}

This difference can not be used as the loss directly, as some exceptional Gaussians should be discarded. One case is that some Gaussians belong to occluded surfaces or objects in the current view. The other case is that non-differentiable regions (such as self-intersection~\cite{mayost_2014_sdfapp}) might lead to a huge difference $\epsilon_{\rm proj}$. These two cases interrupt consistency significantly. Unfortunately, it is difficult to recognize them. Therefore, we use a simple and practical solution, by assuming the difference between the two surfaces should be already close and thresholding the difference with a hyper-parameter $\varepsilon$, as shown in \figref{fig:proj}.
\begin{equation}
\mathcal L_{\rm p}=\frac{1}{N}\sum_{i\in N}\left\{
\begin{aligned}
\epsilon_{\rm proj}^{i} & , & \epsilon_{\rm proj}^{i}\leq\varepsilon, \\
0 & , & \epsilon_{\rm proj}^{i}>\varepsilon.
\end{aligned}
\right.
\end{equation}
Note that we apply this loss after 1K iterations when the coarse geometry is stable so that the assumption holds. 

The theory behind the loss is an approximation of the Eikonal condition, proven in the supplementary. This loss encourages smooth reconstructed surfaces while maintaining the flexibility from Gaussian for geometry details. 

\subsection{Spherical initialization}
\label{ssec:init}

We found that some erroneous geometry emerges at the early stage of training and can hardly be recovered. This indicates that the initialization makes the geometry prone to trapping into local minima. \revise{Since sphere-like initializations revealed their effect in volumetric rendering of SDF~\cite{wang_2021_neus, yariv_2021_volsdf}, we design a spherical initialization of foreground objects tailored for Gaussian splatting to avoid local minima and to encourage optimal geometry.} 
Specifically, we employ the points uniformly sampled on a unit sphere to initialize Gaussians. The design mitigates broken surfaces and overfitting under the training appearance, further improving the geometry robustness.

\section{Experiments}

\mycfigureW{relit_glossy}{relit_glossy_iccv_v2.pdf}{The comparison of relighting results and normal on the Glossy Blender (the $1^{\rm st}$-$4^{\rm th}$ rows) and the TensoIR synthetic dataset (the $5^{\rm th}$, $6^{\rm th}$ rows). Our method provides reasonable normal robustly across materials, thus rendering realistic relighting results that include specular highlights. \revise{The PSNR/SSIM of relighting results under the current view are below images.} More results are in the supplementary.}{0.95}

In this section, we present our training details (\secref{ssec:impl}) and evaluation setup (\secref{ssec:setup}). Then, we compare our method with the representative methods (\secref{ssec:comp}) and conduct ablation studies on our proposed designs (\secref{ssec:ablation}).

\subsection{Implementation details}
\label{ssec:impl}

We optimized our model using Adam optimizer~\cite{kingma_2014_adam} for 30,000 iterations with the loss:
\begin{equation}
    \mathcal{L} = \mathcal{L}_{\rm c} + \lambda_{n}\mathcal{L}_{\rm n} + \lambda_{\rm d}\mathcal{L}_{\rm d} + \lambda_{\gamma}\mathcal{L}_{\gamma} + \lambda_{\rm p}\mathcal{L}_{\rm p} + \lambda_{\rm sm}\mathcal{L}_{\rm sm} + \lambda_{\rm m} \mathcal{L}_{\rm m},
\end{equation}
where $\lambda_{[\cdot]}$ is the corresponding weight of loss $\mathcal{L}_{[\cdot]}$. $\mathcal{L}_{\rm c}$ is the rendering loss between the rendered image and the ground truth. $\mathcal{L}_{\rm n}$ and $\mathcal{L}_{\rm d}$ are the normal consistency and the distortion loss, inheriting from 2DGS. $\mathcal{L}_{\rm sm}$ is a smoothness regularization for PBR attributes, and $\mathcal{L}_{\rm m}$ is the optional mask loss widely adopted by existing relighting methods~\cite{gao_2023_relightablegs, zhang_2021_physg}. $\mathcal{L}_{\gamma}$ and $\mathcal{L}_{\rm p}$ are our median and projection-based loss. More details are in the supplementary.

\subsection{Evaluation setup}
\label{ssec:setup}

\noindent
\textbf{Dataset.} 
We evaluate our method on three synthetic datasets, including TensoIR~\cite{jin_2023_tensoir} synthetic dataset, Glossy Blender dataset~\cite{liu_2023_nero} and Shiny Blender dataset. For real scenes, we select some objects from NeILF++~\cite{zhang_2023_neilfpp}.

\noindent
\textbf{Methods for comparison.} We select 4 representative Gaussian-based inverse rendering methods for comparison, including GShader~\cite{jiang_2024_gaussianshader}, GS-IR~\cite{liang_2024_gsir}, R3DG~\cite{gao_2023_relightablegs} and GS-ROR~\cite{zhu_2024_gsror}. We trained these models based on their public codes and configurations with an extra mask loss for stabilizing training and fair comparison.

\mycfigureW{relit_neilf}{relit_neilfpp.pdf}{The qualitative comparison of relighting results and normal on the NeILF++ dataset. Our method can provide realistic relighting results and detailed normal for real data. More results are in the supplementary.}{0.95}
\mycfigureW{normal_shiny}{normal_shiny_iccv_v3.pdf}{The qualitative comparison of normal on the Shiny Blender dataset. Our method provides globally smooth normal while preserving the sharp details. \revise{The MAE under the current view is below the visualization.} More results are in the supplementary.}{0.95}
\mycfigureW{ablation}{ablation_iccv}{\revise{Ablation of the key components} in our method on the Glossy Blender dataset, including the discretized SDF framework with the median loss, the projection-based consistency loss (Proj.), and spherical initialization (Sph.). (The \textcolor{red}{red} arrow points out the artifacts.)}{0.95}

\noindent
\textbf{Metrics.} We use the peak signal-to-noise ratio (PSNR) and structural similarity index (SSIM)~\cite{wang_2004_ssim}, the learned perceptual image patch similarity (LPIPS)~\cite{zhang_2018_lpips} to measure the relighting quality. Following NeRO and TensoIR, we rescale the relighting images for quantitative evaluation. We compare the mean angular error (MAE) and \revise{the Chamber distance (CD) for geometry quality}. We record the mean training time, the frame per second (FPS), and the memory usage during training on RTX 4090.

\subsection{Comparison with previous methods}
\label{ssec:comp}

\noindent
\textbf{Relighting of synthetic objects.} We evaluate the relighting performance of synthetic objects with various materials on the Glossy Blender and TensoIR synthetic datasets. The quantitative measurements are shown in \tabref{tab:glossy_relit} and \tabref{tab:tir_relit}. The relighting results with normal visualizations are in \figref{fig:relit_glossy}. Due to the robustness of decomposition, our method provides high-quality relighting results and smooth normal. In contrast, GS-IR, R3DG, and GShader overfit the training appearance, leading to erroneous normal. Regarding GS-ROR, it shows higher geometry quality, as it introduces a continuous SDF for regularization. However, the geometry details are lost in some regions, due to the difficulty in balancing two representations during training. 

\begin{table}[tb]
    \centering
\resizebox{\linewidth}{!}{
    \renewcommand{\arraystretch}{1.1}
    \begin{tabular}{lcccccc}
       \hline
            & GS-IR & R3DG & GShader & GS-ROR & Ours  \\
       \hline
       Ball     & 25.79 & 22.44 & 7.03  & \sota{ 0.92 }  & \subsota{ 1.20 }  \\
       Car      & 28.31 & 26.02 & 14.05 & \subsota{11.98} & \sota{ 6.82 }  \\ 
       Coffee   & 15.38 & 13.39 & 14.93 & \subsota{12.24} & \sota{11.42} \\ 
       Helmet   & 25.58 & 19.63 & 9.33  & \subsota{ 4.10 }  & \sota{ 3.90 } \\ 
       Teapot   & 15.35 & 9.21  & 7.17  & \subsota{ 5.88 }  & \sota{ 5.02 } \\ 
       Toaster  & 33.51 & 28.17 & 13.08 & \sota{ 8.24 }  & \subsota{10.51} \\ 
       \hline
       Mean     & 23.99 & 19.81 & 10.93 & \subsota{ 7.23 } & \sota{ 6.48 } \\
       \hline
    \end{tabular}
    }
    \vspace{-2mm}
    \caption{Normal quality with Gaussian-based methods on Shiny Blender dataset in terms of MAE$\downarrow$. Numbers in \sota{red} indicate the best performance and numbers in \subsota{orange} indicate the second best.}
    \vspace{-2mm}
    \label{tab:shiny_mae}
\end{table}
\begin{table}[tb]
    \centering
\resizebox{\linewidth}{!}{
    \renewcommand{\arraystretch}{1.1}
    \begin{tabular}{lcccccc}
       \hline
            & GS-IR & R3DG & GShader & GS-ROR & Ours  \\
       \hline
       Angel     & 0.0177 & 0.0098 & 0.0085 & \subsota{0.0063} & \sota{0.0053}  \\
       Bell      & 0.1153 & 0.0418 & 0.0110 & \subsota{0.0096} & \sota{0.0094}  \\ 
       Cat       & 0.0588 & 0.0339 & 0.0256 & \sota{0.0222} & \subsota{0.0251} \\ 
       Horse     & 0.0196 & 0.0135 & 0.0073 & \subsota{0.0061} & \sota{0.0052} \\ 
       Luyu      & 0.0225 & 0.0168 & 0.0107 & \sota{0.0098} & \subsota{0.0106} \\ 
       Potion    & 0.0623 & 0.0380 & 0.0474 & \subsota{0.0146} & \sota{0.0081}\\ 
       Tbell     & 0.1021 & 0.0500 & 0.0574 & \subsota{0.0270} & \sota{0.0101} \\ 
       Teapot    & 0.0719 & 0.0479 & 0.0340 & \subsota{0.0166} & \sota{0.0115}\\ 
       \hline
       Mean      & 0.0588 & 0.0315 & 0.0253 & \subsota{0.0140} & \sota{0.0107} \\
       \hline
    \end{tabular}
    }
    \vspace{-2mm}
    \caption{\revise{Surface quality with Gaussian-based methods} on Glossy Blender dataset regarding CD$\downarrow$. Numbers in \sota{red} indicate the best performance, and numbers in \subsota{orange} indicate the second best.}
    \vspace{-2mm}
    \label{tab:glossy_cd}
\end{table}
\noindent
\textbf{Relighting of real objects.}
The results of real scenes from NeILF++ are shown in \figref{fig:relit_neilf}. As ground truth is unavailable, we show reference training views and the results under novel light conditions.\footnote{R3DG, GS-IR do not converge on the scenes, thus no results provided.} Our method produces reasonable relighting results with sharp geometry details preserved.

\noindent

\revise{\textbf{Geometry quality.} We compare reconstructed geometry quality in terms of normal MAE in \tabref{tab:shiny_mae} with the visualization in \figref{fig:normal_shiny} and CD in \tabref{tab:glossy_cd}.\footnotemark  Numerically, our model outperforms all Gaussian-based inverse rendering methods on most scenes and average metrics. However, our method fails in scenes with complex inter-reflection (`Toaster', `Luyu', `Cat') but still achieves sub-SOTA performance. Only GS-ROR succeeds in the scenes, at the cost of $5\times$ of GPU memory usage than ours, due to their extra SDF network. Visually, our model \textbf{provides globally smooth normal (teapot in \figref{fig:relit_glossy}) and preserves local details (horse tail in \figref{fig:relit_glossy})}, while GS-ROR  \textit{discards sharp details} and others \textit{fail to reconstruct smooth surfaces.}}
\footnotetext{The choice of quantitative metrics depends on the available ground truth, and we use TSDF fusion~\cite{open3d} for surface extraction to evaluate CD.}

          
        

\begin{table}[tb]
    \centering
    \resizebox{0.99\linewidth}{!}{
    \renewcommand{\arraystretch}{1.2}
    \begin{tabular}{ccc|ccc}
        \hline
        \multicolumn{3}{c|}{Components} & \multicolumn{3}{c}{Scene} \\
       \hline
         \multirow{2}{*}{SDF} & \multirow{2}{*}{Sph.} & \multirow{2}{*}{Proj.}               & Angel       & Horse        & Teapot  \\
                &       &                  & PSNR/SSIM & PSNR/SSIM & PSNR/SSIM  \\
          
       \hline
                  &           &            & 20.23/0.8533 & 20.72/0.8998 & 19.22/0.8674 \\
        \ding{52} &           &            & 21.09/0.8815 & 22.13/0.9120 & 23.26/0.9223 \\
                  &  \ding{52} &           & 21.43/0.8881 & 21.96/0.9102 & 22.97/0.9176 \\
        \ding{52} &  \ding{52} &           & 21.59/0.8910 & 22.87/0.9284 & 23.45/0.9222 \\
        
        \ding{52} &           & \ding{52} & \subsota{21.65/0.8917} & \subsota{23.21/0.9395} & \subsota{24.05/0.9289} \\
        \ding{52} & \ding{52} & \ding{52}  & \sota{22.03/0.8919} & \sota{24.01/0.9481} & \sota{24.19/0.9293} \\ 
       \hline
    \end{tabular}
    }
    \vspace{-2mm}
    \caption{\revise{Ablation study of three key components on the Glossy Blender dataset. ``SDF'' means incorporating the discretized SDF with the median loss, ``Proj.'' means the projection-based consistency loss, and ``Sph.'' means the spherical initialization.}}\vspace{-2mm}
    \label{tab:ab}
\end{table}

\subsection{Ablation study}
\label{ssec:ablation}

We conduct ablation studies of our main components on the Glossy Blender dataset, and \revise{the ablations of detailed choices (\eg losses, threshold) are in the supplementary.} The quantitative results are in the \tabref{tab:ab}. 
The performance of the model improves consistently when employing new components.
We start from the baseline described in \secref{ssec:sdf_gs}. As shown in the $1^{\rm st}$ column of \figref{fig:ablation}, this baseline overfits the training appearance and fails to decompose geometry and material for relighting. 

\noindent
\textbf{Discretized SDF with the median loss.} Then, we incorporate the SDF-to-opacity transformation and the median loss that promotes the Gaussians converging to the opaque surface. As shown in the $2^{\rm nd}$ column of \figref{fig:ablation}. The variant can decompose the material and geometry more robustly, thus improving the relighting quality. However, due to insufficient constraints, we still observe erroneous normal.

\noindent
\textbf{Projection-based consistency loss.} \revise{Due to this loss is coupled with SDF, it cannot be ablated individually.} Applying the projection-based consistency loss leads to more reasonable normal. 
However, erroneous geometry caused by misleading initialization still exists, as in the $5^{\rm th}$ row of \figref{fig:ablation}.

\noindent
\textbf{Spherical initialization.} \revise{The spherical initialization leads to performance increase with or without SDF designs (the $3^{\rm rd}, 4^{\rm th}, 6^{\rm th}$ columns of \figref{fig:ablation}). However, applying it individually cannot ensure artifact-free results.}


\subsection{Discussion and limitations}

In this work, we focus on improving the relighting quality by introducing SDF prior and leave mesh extraction with BRDF parameters for future work. Besides, we consider the direct lighting only for efficiency, so our method may fail for objects with complex occlusion. An extra indirect illumination term~\cite{gao_2023_relightablegs,liu_2023_nero} will further benefit our approach.


\section{Conclusion}

In this paper, we presented our Gaussian splatting framework for real-time object relighting, which regularizes the geometry from Gaussian primitives with a discretized SDF. In this framework, we proposed the median loss to facilitate the convergence and the projection-based consistency loss to constrain the discrete samples of SDF, along with the spherical initialization to avoid local minima. Therefore, our method benefits from the flexibility of GS and the robustness of SDF, thus outperforming the existing Gaussian-based inverse rendering methods in terms of relighting and decomposition quality, while maintaining simplicity and efficiency. In future work, extending our framework to unbounded scenes is a promising direction. 

\small{\noindent\textbf{Acknowledgment:}
We thank the reviewers for the valuable comments. This work has been partially supported by the National Science and Technology Major Project under grant No. 2022ZD0116305, National Natural Science Foundation of China (NSFC) under grant No. 62172220, and the NSFC under Grant Nos. 62361166670 and U24A20330. Computation is supported by the Supercomputing Center of Nankai University.}
{
    \small
    \bibliographystyle{ieeenat_fullname}
    \bibliography{main}

\begin{thebibliography}{44}
\providecommand{\natexlab}[1]{#1}
\providecommand{\url}[1]{\texttt{#1}}
\expandafter\ifx\csname urlstyle\endcsname\relax
  \providecommand{\doi}[1]{doi: #1}\else
  \providecommand{\doi}{doi: \begingroup \urlstyle{rm}\Url}\fi

\bibitem[Baorui et~al.(2021)Baorui, Zhizhong, Yu-Shen, and Matthias]{ma_2021_np}
Ma Baorui, Han Zhizhong, Liu Yu-Shen, and Zwicker Matthias.
\newblock Neural-{P}ull: Learning signed distance functions from point clouds by learning to pull space onto surfaces.
\newblock In \emph{ICML}, 2021.

\bibitem[Bi et~al.(2020)Bi, Xu, Srinivasan, Mildenhall, Sunkavalli, Hašan, Hold-Geoffroy, Kriegman, and Ramamoorthi]{bi_2020_neural}
Sai Bi, Zexiang Xu, Pratul Srinivasan, Ben Mildenhall, Kalyan Sunkavalli, Miloš Hašan, Yannick Hold-Geoffroy, David Kriegman, and Ravi Ramamoorthi.
\newblock Neural reflectance fields for appearance acquisition.
\newblock \emph{arXiv preprint arXiv:2008.03824}, 2020.

\bibitem[Boss et~al.(2021{\natexlab{a}})Boss, Braun, Jampani, Barron, Liu, and Lensch]{boss_2021_nerd}
Mark Boss, Raphael Braun, Varun Jampani, Jonathan~T. Barron, Ce Liu, and Hendrik~P.A. Lensch.
\newblock Ne{RD}: Neural reflectance decomposition from image collections.
\newblock In \emph{ICCV}, pages 12684--12694, New York, NY, USA, 2021{\natexlab{a}}. IEEE.

\bibitem[Boss et~al.(2021{\natexlab{b}})Boss, Jampani, Braun, Liu, Barron, and Lensch]{boss_2021_neuralpil}
Mark Boss, Varun Jampani, Raphael Braun, Ce Liu, Jonathan Barron, and Hendrik~PA Lensch.
\newblock Neural-{PIL}: Neural pre-integrated lighting for reflectance decomposition.
\newblock In \emph{NeurIPS}, pages 10691--10704, Red Hook, NY, USA, 2021{\natexlab{b}}. Curran Associates, Inc.

\bibitem[Chen et~al.(2023)Chen, Li, and Lee]{chen_2023_neusg}
Hanlin Chen, Chen Li, and Gim~Hee Lee.
\newblock Neu{SG}: Neural implicit surface reconstruction with 3{D} {G}aussian splatting guidance.
\newblock \emph{arXiv preprint arXiv:2312.00846}, 2023.

\bibitem[Gao et~al.(2024)Gao, Gu, Lin, Li, Zhu, Cao, Zhang, and Yao]{gao_2023_relightablegs}
Jian Gao, Chun Gu, Youtian Lin, Zhihao Li, Hao Zhu, Xun Cao, Li Zhang, and Yao Yao.
\newblock Relightable 3{D} {G}aussians: Realistic point cloud relighting with {BRDF} decomposition and ray tracing.
\newblock In \emph{ECCV}, 2024.

\bibitem[Gropp et~al.(2020)Gropp, Yariv, Haim, Atzmon, and Lipman]{gropp_2020_eik}
Amos Gropp, Lior Yariv, Niv Haim, Matan Atzmon, and Yaron Lipman.
\newblock Implicit geometric regularization for learning shapes.
\newblock In \emph{ICML}. JMLR.org, 2020.

\bibitem[Huang et~al.(2024)Huang, Yu, Chen, Geiger, and Gao]{huang_2024_2dgs}
Binbin Huang, Zehao Yu, Anpei Chen, Andreas Geiger, and Shenghua Gao.
\newblock 2{D} {G}aussian splatting for geometrically accurate radiance fields.
\newblock In \emph{SIGGRAPH}, New York, NY, USA, 2024. Association for Computing Machinery.

\bibitem[Jiang et~al.(2024)Jiang, Tu, Liu, Gao, Long, Wang, and Ma]{jiang_2024_gaussianshader}
Yingwenqi Jiang, Jiadong Tu, Yuan Liu, Xifeng Gao, Xiaoxiao Long, Wenping Wang, and Yuexin Ma.
\newblock {G}aussian{S}hader: 3{D} {G}aussian splatting with shading functions for reflective surfaces.
\newblock In \emph{CVPR}, pages 5322--5332, New York, NY, USA, 2024. IEEE.

\bibitem[Jin et~al.(2023)Jin, Liu, Xu, Zhang, Han, Bi, Zhou, Xu, and Su]{jin_2023_tensoir}
Haian Jin, Isabella Liu, Peijia Xu, Xiaoshuai Zhang, Songfang Han, Sai Bi, Xiaowei Zhou, Zexiang Xu, and Hao Su.
\newblock Tenso{IR}: Tensorial inverse rendering.
\newblock In \emph{CVPR}, pages 165--174, New York, NY, USA, 2023. IEEE.

\bibitem[Karis(2013)]{karis_2013_splitsum}
Brian Karis.
\newblock Real shading in unreal engine 4.
\newblock \emph{Proc. Physically Based Shading Theory Practice}, 2013.

\bibitem[Kerbl et~al.(2023)Kerbl, Kopanas, Leimk{\"u}hler, and Drettakis]{kerbl_2023_3dgs}
Bernhard Kerbl, Georgios Kopanas, Thomas Leimk{\"u}hler, and George Drettakis.
\newblock 3{D} {G}aussian splatting for real-time radiance field rendering.
\newblock \emph{ACM TOG}, 42\penalty0 (4):\penalty0 139--1, 2023.

\bibitem[Kingma and Ba(2017)]{kingma_2014_adam}
Diederik~P. Kingma and Jimmy Ba.
\newblock Adam: A method for stochastic optimization.
\newblock \emph{arXiv preprint arXiv:1412.6980}, 2017.

\bibitem[Li et~al.(2024)Li, Wang, Zhang, and Wang]{li_2024_tensosdf}
Jia Li, Lu Wang, Lei Zhang, and Beibei Wang.
\newblock Tenso{SDF}: Roughness-aware tensorial representation for robust geometry and material reconstruction.
\newblock \emph{ACM TOG}, 43\penalty0 (4), 2024.

\bibitem[Li et~al.(2023)Li, M\"uller, Evans, Taylor, Unberath, Liu, and Lin]{li_2023_neuralangelo}
Zhaoshuo Li, Thomas M\"uller, Alex Evans, Russell~H. Taylor, Mathias Unberath, Ming-Yu Liu, and Chen-Hsuan Lin.
\newblock Neuralangelo: High-fidelity neural surface reconstruction.
\newblock In \emph{CVPR}, pages 8456--8465, New York, NY, USA, 2023. IEEE.

\bibitem[Liang et~al.(2024)Liang, Zhang, Feng, Shan, and Jia]{liang_2024_gsir}
Zhihao Liang, Qi Zhang, Ying Feng, Ying Shan, and Kui Jia.
\newblock G{S}-{IR}: 3{D} {G}aussian splatting for inverse rendering.
\newblock In \emph{CVPR}, pages 21644--21653, New York, NY, USA, 2024. IEEE.

\bibitem[Liu et~al.(2023)Liu, Wang, Lin, Long, Wang, Liu, Komura, and Wang]{liu_2023_nero}
Yuan Liu, Peng Wang, Cheng Lin, Xiaoxiao Long, Jiepeng Wang, Lingjie Liu, Taku Komura, and Wenping Wang.
\newblock Ne{RO}: Neural geometry and {BRDF} reconstruction of reflective objects from multiview images.
\newblock \emph{ACM TOG}, 42\penalty0 (4), 2023.

\bibitem[Lyu et~al.(2024)Lyu, Sun, Huang, Wu, Yang, Chen, Pang, and Qi]{yu_2024_3dgsr}
Xiaoyang Lyu, Yang-Tian Sun, Yi-Hua Huang, Xiuzhe Wu, Ziyi Yang, Yilun Chen, Jiangmiao Pang, and Xiaojuan Qi.
\newblock 3{DGSR}: Implicit surface reconstruction with 3{D} {G}aussian splatting.
\newblock \emph{ACM TOG}, 43\penalty0 (6), 2024.

\bibitem[Mayost(2014)]{mayost_2014_sdfapp}
Daniel Mayost.
\newblock Applications of the signed distance function to surface geometry.
\newblock University of Toronto, 2014.

\bibitem[McAuley et~al.(2013)McAuley, Hill, Martinez, Villemin, Pettineo, Lazarov, Neubelt, Karis, Hery, Hoffman, and Zap~Andersson]{burley_2012_physically}
Stephen McAuley, Stephen Hill, Adam Martinez, Ryusuke Villemin, Matt Pettineo, Dimitar Lazarov, David Neubelt, Brian Karis, Christophe Hery, Naty Hoffman, and Hakan Zap~Andersson.
\newblock Physically based shading in theory and practice.
\newblock In \emph{ACM SIGGRAPH 2013 Courses}, New York, NY, USA, 2013. Association for Computing Machinery.

\bibitem[Mildenhall et~al.(2020)Mildenhall, Srinivasan, Tancik, Barron, Ramamoorthi, and Ng]{mildenhall_2020_nerf}
Ben Mildenhall, Pratul~P. Srinivasan, Matthew Tancik, Jonathan~T. Barron, Ravi Ramamoorthi, and Ren Ng.
\newblock Ne{RF}: Representing scenes as neural radiance fields for view synthesis.
\newblock In \emph{ECCV}, pages 405--421, Berlin, Heidelberg, 2020. Springer.

\bibitem[Rosu and Behnke(2023)]{rosu_2023_permutosdf}
Radu~Alexandru Rosu and Sven Behnke.
\newblock Permuto{SDF}: Fast multi-view reconstruction with implicit surfaces using permutohedral lattices.
\newblock In \emph{CVPR}, pages 8466--8475, New York, NY, USA, 2023. IEEE.

\bibitem[Shi et~al.(2023)Shi, Wu, Wu, Liu, Zhao, Feng, Liu, Zhang, Zhang, Zhou, Ding, and Wang]{shi_2023_gir}
Yahao Shi, Yanmin Wu, Chenming Wu, Xing Liu, Chen Zhao, Haocheng Feng, Jingtuo Liu, Liangjun Zhang, Jian Zhang, Bin Zhou, Errui Ding, and Jingdong Wang.
\newblock G{IR}: 3{D} {G}aussian inverse rendering for relightable scene factorization.
\newblock \emph{arXiv preprint arXiv:2312.05133}, 2023.

\bibitem[Srinivasan et~al.(2021)Srinivasan, Deng, Zhang, Tancik, Mildenhall, and Barron]{srinivasa_2021_nerv}
Pratul~P. Srinivasan, Boyang Deng, Xiuming Zhang, Matthew Tancik, Ben Mildenhall, and Jonathan~T. Barron.
\newblock Ne{RV}: Neural reflectance and visibility fields for relighting and view synthesis.
\newblock In \emph{CVPR}, pages 7495--7504, New York, NY, USA, 2021. IEEE.

\bibitem[Sun et~al.(2023)Sun, Cai, Li, Yan, Zhang, Marshall, Huang, Zhao, and Dong]{sun_2024_neuralpbir}
Cheng Sun, Guangyan Cai, Zhengqin Li, Kai Yan, Cheng Zhang, Carl Marshall, Jia-Bin Huang, Shuang Zhao, and Zhao Dong.
\newblock Neural-{PBIR} reconstruction of shape, material, and illumination.
\newblock In \emph{ICCV}, pages 18046--18056, New York, NY, USA, 2023. IEEE.

\bibitem[Verbin et~al.(2022)Verbin, Hedman, Mildenhall, Zickler, Barron, and Srinivasan]{verbin_2022_refnerf}
Dor Verbin, Peter Hedman, Ben Mildenhall, Todd Zickler, Jonathan~T. Barron, and Pratul~P. Srinivasan.
\newblock Ref-{N}e{RF}: Structured view-dependent appearance for neural radiance fields.
\newblock In \emph{CVPR}, pages 5491--5500, New York, NY, USA, 2022. IEEE.

\bibitem[Wang et~al.(2024)Wang, Hu, Zhu, and Lau]{wang_2024_inverse}
Haoyuan Wang, Wenbo Hu, Lei Zhu, and Rynson~W.H. Lau.
\newblock Inverse rendering of glossy objects via the neural plenoptic function and radiance fields.
\newblock In \emph{CVPR}, pages 19999--20008, New York, NY, USA, 2024. IEEE.

\bibitem[Wang et~al.(2021)Wang, Liu, Liu, Theobalt, Komura, and Wang]{wang_2021_neus}
Peng Wang, Lingjie Liu, Yuan Liu, Christian Theobalt, Taku Komura, and Wenping Wang.
\newblock Neu{S}: Learning neural implicit surfaces by volume rendering for multi-view reconstruction.
\newblock In \emph{NeurIPS}, pages 27171--27183, Red Hook, NY, USA, 2021. Curran Associates, Inc.

\bibitem[Wang et~al.(2023{\natexlab{a}})Wang, Han, Habermann, Daniilidis, Theobalt, and Liu]{wang2023neus2}
Yiming Wang, Qin Han, Marc Habermann, Kostas Daniilidis, Christian Theobalt, and Lingjie Liu.
\newblock Neu{S}2: Fast learning of neural implicit surfaces for multi-view reconstruction.
\newblock In \emph{ICCV}, pages 3295--3306, New York, NY, USA, 2023{\natexlab{a}}. IEEE.

\bibitem[Wang et~al.(2004)Wang, Bovik, Sheikh, and Simoncelli]{wang_2004_ssim}
Zhou Wang, A.C. Bovik, H.R. Sheikh, and E.P. Simoncelli.
\newblock Image quality assessment: from error visibility to structural similarity.
\newblock \emph{IEEE TIP}, 13\penalty0 (4):\penalty0 600--612, 2004.

\bibitem[Wang et~al.(2023{\natexlab{b}})Wang, Zhang, Xu, Zhang, Wang, Chen, Xin, Wang, and Tu]{wang_2023_neuralsingular}
Zixiong Wang, Yunxiao Zhang, Rui Xu, Fan Zhang, Peng-Shuai Wang, Shuangmin Chen, Shiqing Xin, Wenping Wang, and Changhe Tu.
\newblock Neural-singular-hessian: Implicit neural representation of unoriented point clouds by enforcing singular hessian.
\newblock \emph{ACM TOG}, 42\penalty0 (6), 2023{\natexlab{b}}.

\bibitem[Wu et~al.(2024)Wu, Sun, Lai, Ma, Kobbelt, and Gao]{wu_2024_deferredgs}
Tong Wu, Jia-Mu Sun, Yu-Kun Lai, Yuewen Ma, Leif Kobbelt, and Lin Gao.
\newblock Deferred{GS}: Decoupled and editable {G}aussian splatting with deferred shading.
\newblock \emph{arXiv preprint arXiv:2404.09412}, 2024.

\bibitem[Yariv et~al.(2021)Yariv, Gu, Kasten, and Lipman]{yariv_2021_volsdf}
Lior Yariv, Jiatao Gu, Yoni Kasten, and Yaron Lipman.
\newblock Volume rendering of neural implicit surfaces.
\newblock In \emph{NeurIPS}, pages 4805--4815, Red Hook, NY, USA, 2021. Curran Associates, Inc.

\bibitem[Ye et~al.(2024)Ye, Hou, and Zhou]{ye_2024_3dgsdr}
Keyang Ye, Qiming Hou, and Kun Zhou.
\newblock 3{D} {G}aussian splatting with deferred reflection.
\newblock In \emph{SIGGRAPH}, New York, NY, USA, 2024. Association for Computing Machinery.

\bibitem[Yu et~al.(2024)Yu, Lu, Xu, Jiang, Xiangli, and Dai]{yu_2024_gsdf}
Mulin Yu, Tao Lu, Linning Xu, Lihan Jiang, Yuanbo Xiangli, and Bo Dai.
\newblock G{SDF}: 3{DGS} meets {SDF} for improved rendering and reconstruction.
\newblock In \emph{NeurIPS}, 2024.

\bibitem[Zhang et~al.(2022{\natexlab{a}})Zhang, Yao, Li, Fang, McKinnon, Tsin, and Quan]{zhang_2022_hes}
Jingyang Zhang, Yao Yao, Shiwei Li, Tian Fang, David McKinnon, Yanghai Tsin, and Long Quan.
\newblock Critical regularizations for neural surface reconstruction in the wild.
\newblock In \emph{CVPR}, pages 6270--6279, New York, NY, USA, 2022{\natexlab{a}}. IEEE.

\bibitem[Zhang et~al.(2023)Zhang, Yao, Li, Liu, Fang, McKinnon, Tsin, and Quan]{zhang_2023_neilfpp}
Jingyang Zhang, Yao Yao, Shiwei Li, Jingbo Liu, Tian Fang, David McKinnon, Yanghai Tsin, and Long Quan.
\newblock Ne{ILF}++: Inter-reflectable light fields for geometry and material estimation.
\newblock In \emph{ICCV}, pages 3601--3610, New York, NY, USA, 2023. IEEE.

\bibitem[Zhang et~al.(2021)Zhang, Luan, Wang, Bala, and Snavely]{zhang_2021_physg}
Kai Zhang, Fujun Luan, Qianqian Wang, Kavita Bala, and Noah Snavely.
\newblock Phy{SG}: Inverse rendering with spherical {G}aussians for physics-based material editing and relighting.
\newblock In \emph{CVPR}, pages 5453--5462, New York, NY, USA, 2021. IEEE.

\bibitem[Zhang et~al.(2022{\natexlab{b}})Zhang, Luan, Li, and Snavely]{zhang_iron_2022}
Kai Zhang, Fujun Luan, Zhengqi Li, and Noah Snavely.
\newblock I{RON}: Inverse rendering by optimizing neural sdfs and materials from photometric images.
\newblock In \emph{CVPR}, pages 5565--5574, New York, NY, USA, 2022{\natexlab{b}}. IEEE.

\bibitem[Zhang et~al.(2018)Zhang, Isola, Efros, Shechtman, and Wang]{zhang_2018_lpips}
Richard Zhang, Phillip Isola, Alexei~A. Efros, Eli Shechtman, and Oliver Wang.
\newblock The unreasonable effectiveness of deep features as a perceptual metric.
\newblock In \emph{CVPR}, New York, NY, USA, 2018. IEEE.

\bibitem[Zhang et~al.(2024)Zhang, Liu, and Han]{zhang_2024_gspull}
Wenyuan Zhang, Yu-Shen Liu, and Zhizhong Han.
\newblock Neural signed distance function inference through splatting 3{D} {G}aussians pulled on zero-level set.
\newblock In \emph{NeurIPS}, 2024.

\bibitem[Zhang et~al.(2022{\natexlab{c}})Zhang, Sun, He, Fu, Jia, and Zhou]{zhang_2022_mii}
Yuanqing Zhang, Jiaming Sun, Xingyi He, Huan Fu, Rongfei Jia, and Xiaowei Zhou.
\newblock Modeling indirect illumination for inverse rendering.
\newblock In \emph{CVPR}, pages 18643--18652, New York, NY, USA, 2022{\natexlab{c}}. IEEE.

\bibitem[Zhou et~al.(2018)Zhou, Park, and Koltun]{open3d}
Qian-Yi Zhou, Jaesik Park, and Vladlen Koltun.
\newblock Open3{D}: A modern library for 3d data processing.
\newblock \emph{arXiv preprint arXiv:1801.09847}, 2018.

\bibitem[Zhu et~al.(2024)Zhu, Wang, and Yang]{zhu_2024_gsror}
Zuo-Liang Zhu, Beibei Wang, and Jian Yang.
\newblock G{S}-{ROR}: 3{D} {G}aussian splatting for reflective object relighting via sdf priors.
\newblock \emph{arXiv preprint arXiv:2406.18544}, 2024.

\end{thebibliography}
}


\section*{Supplementary material}

# Gaussian Splatting with Discretized SDF for Relightable Assets

## Supplementary Material

## 1. Model details

In this section, we complete the details of our model, including the losses, hyperparameters, shading model, and lighting model.

### 1.1. Loss details

We introduce the definition and weights of losses we used in the main paper. The color loss is defined as in 3DGS [4]

$$\mathcal{L}_c = \lambda ||C_{\text{gs}} - C_{\text{gt}}||_1 + (1-\lambda)(1 - \text{SSIM}(C_{\text{gs}}, C_{\text{gt}})), \quad (1)$$

where $C_{\text{gs}}$ is the rendered color from Gaussians and $C_{\text{gt}}$ is the ground truth color. The coefficient $\lambda$ is set to $0.8$. The supervision $L_n$ from normal is defined as

$$\mathcal{L}_n = ||\hat{n} - n||^2, \quad (2)$$

where $\hat{n}, n$ are the normal from depth and the normal from Gaussians, respectively. We follow the 2DGS and apply the distortion loss as

$$\mathcal{L}_d = \sum_{i,j} w_i w_j |D_i - D_j|, \quad (3)$$

where $w_i$ is the blending weight of $i$-th Gaussian and $D_i$ is the intersection depth for the current pixel. The smoothness loss for BRDF parameters (*i.e.,* albedo **a**, roughness **r**, metallicity **m**) is defined as

$$\mathcal{L}_{\text{sm}} = ||\nabla p|| \exp^{-||\nabla C_{\text{gt}}||}, (p \in \{\mathbf{a}, \mathbf{r}, \mathbf{m}\}) \quad (4)$$

where $\nabla$ is the gradient operator and $C_{\text{gt}}$ is the color of ground-truth image. The mask loss is the binary cross entropy between predicted and ground-truth masks. The corresponding weights for the above losses are shown in Tab. 1.

The learning rate of Gaussian attributes follows Gaussian Shader [1], except for the rate of SDF value (set to $0.05$). The details of losses are present in the supplementary. The entire training takes about 1 hour on an RTX 4090.

### 1.2. Hyperparameters

As the median loss is used to promote convergence, we remove this loss when the SDF median $s_m < 0.2$, which indicates the transformation has been converged. Besides, the relaxing threshold $\varepsilon$ for the projection-based consistency loss is set to $0.05$. The number $N$ for the spherical initialization is $10^6$.

| Loss | $\mathcal{L}_n$ | $\mathcal{L}_d$ | $\mathcal{L}_\gamma$ | $\mathcal{L}_p$ | $\mathcal{L}_{\text{sm}}$ | $\mathcal{L}_m$ |
|---|---|---|---|---|---|---|
| Weight | 0.2 | 2000 | 1 | 10 | 0.05 | 0.2 |

Table 1. The weights of losses present in the paper.

### 1.3. Split-sum approximation

A typical rendering equation needs to compute an integral on the upper hemisphere involving incident light, view direction, normal, and Bidirectional Reflectance Distribution Function (BRDF):

$$c(\omega_o) = \int_\Omega L_i(\omega_i) f(\omega_i, \omega_o)(\omega_i \cdot n) d\omega_i, \quad (5)$$

where $L_i(\omega_i)$ denotes the light from incident direction $\omega_i$, $f(\omega_i, \omega_o)$ is the BRDF with respect to the incident direction $\omega_i$ and outgoing direction $\omega_o$, and $n$ denotes the surface normal. In practice, the Disney Principled BRDF [7] is the most widely used BRDF, consisting of a diffuse lobe and a specular lobe

$$f(\omega_i, \omega_o) = \underbrace{(1-\mathbf{m})\frac{\mathbf{a}}{\pi}}_{\text{diffuse}} + \underbrace{\frac{DFG}{4(\omega_i \cdot n)(\omega_o \cdot n)}}_{\text{specular}}, \quad (6)$$

where $D$ is the normal distribution function, $F$ is the Fresnel term, $G$ is the geometry term, $\mathbf{a}$ is the albedo, and $\mathbf{m}$ is the metallicity. However, the integral is expensive to evaluate, and the split-sum technique [3] is an alternative widely used in real-time rendering. The specular term is approximated as

$$c_{\text{specular}} = L_{\text{specular}} \cdot ((1-\mathbf{m}) \times 0.04 + \mathbf{m} \times \mathbf{a}) \times F_1 + F_2, \quad (7)$$

where $F_1, F_2$ are two scalars from a pre-computed table. The diffuse term is computed as

$$c_{\text{diffuse}} = \frac{\mathbf{a}(1-\mathbf{m})}{\pi} \cdot L_{\text{diffuse}}. \quad (8)$$

$L_{\text{diffuse}}$ and $L_{\text{specular}}$ are directed queried for a pre-filtered environment map or neural light representations. The rendered color $c(\omega_o)$ is computed as

$$c(\omega_o) = c_{\text{diffuse}} + c_{\text{specular}}. \quad (9)$$

### 1.4. Lighting model

Considering the efficiency of the light query, we use a simple differential environment map and do not model the indirect illumination and occlusion. The environment light is in the cube map format, whose resolution is $6 \times 512 \times 512$.

|         | MII | TensoIR | TensoSDF | NeRO | Ours |
|---------|-----|---------|----------|------|------|
|         | PSNR / SSIM / LPIPS | PSNR / SSIM / LPIPS | PSNR / SSIM / LPIPS | PSNR / SSIM / LPIPS | PSNR / SSIM / LPIPS |
| Angel   | 16.24 / 0.8236 / 0.1404 | 10.24 / 0.2238 / 0.2739 | 20.40 / 0.8969 / 0.0871 | 16.21 / 0.7819 / 0.1923 | 22.03 / 0.8919 / 0.0819 |
| Bell    | 17.41 / 0.8594 / 0.1534 | 10.11 / 0.1018 / 0.2806 | 29.91 / 0.9767 / 0.0263 | 31.19 / 0.9794 / 0.0189 | 24.67 / 0.9280 / 0.0842 |
| Cat     | 17.68 / 0.8521 / 0.1429 | 9.10 / 0.1644 / 0.2146 | 26.12 / 0.9354 / 0.0675 | 28.42 / 0.9579 / 0.0455 | 26.48 / 0.9374 / 0.0661 |
| Horse   | 20.98 / 0.8997 / 0.0713 | 10.42 / 0.1931 / 0.2913 | 27.18 / 0.9567 / 0.0318 | 25.56 / 0.9437 / 0.0410 | 24.01 / 0.9481 / 0.0351 |
| Luyu    | 17.89 / 0.8050 / 0.1393 | 8.27 / 0.2375 / 0.2463 | 19.91 / 0.8825 / 0.0807 | 26.22 / 0.9092 / 0.0696 | 23.80 / 0.9017 / 0.0699 |
| Potion  | 17.13 / 0.8094 / 0.1747 | 6.21 / 0.0846 / 0.2954 | 27.71 / 0.9422 / 0.0759 | 30.14 / 0.9561 / 0.0623 | 27.31 / 0.9280 / 0.0982 |
| Tbell   | 16.54 / 0.8262 / 0.1938 | 7.47 / 0.1609 / 0.2786 | 23.33 / 0.9404 / 0.0543 | 25.45 / 0.9607 / 0.0407 | 23.66 / 0.9191 / 0.0981 |
| Teapot  | 16.71 / 0.8546 / 0.1426 | 9.96 / 0.2093 / 0.2030 | 25.16 / 0.9482 / 0.0485 | 29.87 / 0.9755 / 0.0193 | 24.19 / 0.9293 / 0.0760 |
| Mean    | 17.57 / 0.8413 / 0.1448 | 8.97 / 0.1719 / 0.2605 | 24.97 / 0.9349 / 0.0590 | 26.63 / 0.9331 / 0.0612 | 24.52 / 0.9229 / 0.0762 |
| Training | 4h | 5h | 6h | 12h | 1h |
| Ren. FPS | 1/30 | 1/60 | 1/4 | 1/4 | 143 |
| Memory  | 12G | 23G | 20G | 8G | 4G |

Table 2. The quantitative comparison with NeRF-based methods on the Glossy Blender dataset in terms of PSNR↑, SSIM↑, and LPIPS↓. Numbers in red indicate the best performance, numbers in orange indicate the second best, and numbers in yellow indicate the third best. Although our method has a performance gap compared to NeRO and TensoSDF, our method is more efficient (at most **50% memory usage and 17% training time**).

|         | MII | NeRO | TensoSDF | TensoIR | Ours |
|---------|-----|------|----------|---------|------|
|         | PSNR / SSIM / LPIPS | PSNR / SSIM / LPIPS | PSNR / SSIM / LPIPS | PSNR / SSIM / LPIPS | PSNR / SSIM / LPIPS |
| Armad.  | 26.85 / 0.9441 / 0.0692 | 23.02 / 0.9335 / 0.0644 | 23.02 / 0.9355 / 0.0578 | 34.51 / 0.9754 / 0.0368 | 31.05 / 0.9621 / 0.0536 |
| Ficus   | 20.65 / 0.9068 / 0.0728 | 27.43 / 0.9404 / 0.0677 | 28.53 / 0.9499 / 0.0533 | 24.32 / 0.9465 / 0.0543 | 27.85 / 0.9639 / 0.0390 |
| Hotdog  | 22.65 / 0.9011 / 0.0893 | 20.45 / 0.9262 / 0.0888 | 20.47 / 0.9241 / 0.0906 | 27.92 / 0.9324 / 0.0833 | 26.23 / 0.9360 / 0.0746 |
| Lego    | 23.20 / 0.8643 / 0.1715 | 17.76 / 0.8577 / 0.1228 | 17.92 / 0.8670 / 0.1088 | 27.61 / 0.9253 / 0.0702 | 25.81 / 0.9087 / 0.0791 |
| Mean    | 23.34 / 0.9041 / 0.1007 | 22.17 / 0.9145 / 0.0859 | 22.48 / 0.9191 / 0.0776 | 28.59 / 0.9449 / 0.0612 | 27.78 / 0.9427 / 0.0626 |

Table 3. The quantitative comparison with NeRF-based methods on the TensoIR Synthetic dataset in terms of PSNR↑, SSIM↑, and LPIPS↓. Numbers in red indicate the best performance, numbers in orange indicate the second best, and numbers in yellow indicate the third best.

## 2. Proof of the projection-based loss

This part reveals how our projection-based consistency loss approximates the Eikonal loss.

**Proposition 1.** *Consider a differentiable function $f$ in the region $\Omega$, for $\forall x_1 \in \Omega, f(x_1) \neq 0$ and its projection point $x_0 = x_1 - \nabla f(x_1)/|\nabla f(x_1)| \cdot f(x_1)$, if we guarantee $x_0$ on the surface ($f(x_0) = 0$), the Eikonal condition $|\nabla f(x)| = 1$ can be approximated at least in a narrow thin-shell space surrounding the surface.*

*Proof.* The Taylor expansion of $f$ at position $x_1$ is

$$f(x) = f(x_1) + \nabla f(x_1)(x - x_1) + O((x - x_1)^2) \quad (10)$$

where $\nabla f(x_1)$ means the gradient of $f$ at $x_1$ and $O((x - x_1)^2)$ means the higher-order term of $f$. Then substitute $x_0$ into $f(x)$:

$$x_0 - x_1 = -f(x_1) \cdot \nabla f(x_1)/|\nabla f(x_1)|,$$

$$f(x_0) = f(x_1) - f(x_1)\frac{\nabla f(x_1)^2}{|\nabla f(x_1)|} + O((x_0 - x_1)^2) \quad (11)$$

$$f(x_1)(1 - |\nabla f(x_1)|) + O((x_0 - x_1)^2) = 0$$

|        | GS-ROR | Ours w/o mask | Ours w/ mask |
|--------|--------|---------------|--------------|
|        | PSNR / SSIM | PSNR / SSIM | PSNR / SSIM |
| Angel  | 20.81 / 0.8775 | 21.03 / 0.8842 | 22.03 / 0.8919 |
| Horse  | 23.31 / 0.9376 | 23.35 / 0.9406 | 24.01 / 0.9481 |
| Teapot | 21.17 / 0.8932 | 22.57 / 0.9087 | 24.19 / 0.9293 |

Table 4. The impact of the mask loss on our method.

which means $|\nabla f(x)| = 1$ is approximated for each $x_1$ where $x_0, x_1$ is close enough to make the high-order term $O((x_0 - x_1)^2)$ negligible. Therefore, we can guarantee the Eikonal condition near the surface where Gaussians exist.

## 3. More results

In this section, we present the ablation of detailed choices in our model and more results of our method. Additionally, we select some representative NeRF-based methods for further validation, including MII [10], TensoIR [2], TensoSDF [5], and NeRO [6].

| $\varepsilon$ | 0.01 | 0.05 | 0.1 | 0.2 |
|---|---|---|---|---|
| | PSNR / SSIM | PSNR / SSIM | PSNR / SSIM | PSNR / SSIM |
| Angel | 21.78 / 0.8860 | 21.91 / 0.8880 | 22.03 / 0.8919 | 21.54 / 0.8769 |
| Horse | 22.10 / 0.9214 | 22.51 / 0.9317 | 24.01 / 0.9481 | 20.12 / 0.9220 |
| Teapot | 23.28 / 0.9190 | 23.84 / 0.9232 | 24.19 / 0.9293 | 23.37 / 0.9200 |

Table 5. Ablation study on the threshold $\varepsilon$ in our method.

| Iterations | 7K | 15K | 30K |
|---|---|---|---|
| | PSNR / SSIM | PSNR / SSIM | PSNR / SSIM |
| No reg. | 17.96 / 0.8881 | 22.13 / 0.9175 | 22.82 / 0.9330 |
| $\gamma = \gamma_\mathrm{m}$ | 18.85 / 0.9032 | 22.46 / 0.9239 | 22.91 / 0.9332 |
| $\mathcal{L}_\gamma$ | 22.20 / 0.9221 | 23.64 / 0.9319 | 24.19 / 0.9481 |

Table 6. Ablation study on the median loss $\mathcal{L}_\gamma$ in our method. "No reg." means imposing no regularization on the transformation.

| | GS-IR | GShader | R3DG | GS-ROR | Ours |
|---|---|---|---|---|---|
| | PSNR / CD | PSNR / CD | PSNR / CD | PSNR / CD | PSNR / CD |
| Ball | 18.30/—— | 30.40/—— | 21.39/—— | 35.50/—— | 34.64/—— |
| Car | 25.30/0.61 | 28.39/0.26 | 26.59/0.23 | 30.52/0.17 | 30.55/0.14 |
| Coffee | 30.72/1.22 | 30.79/1.05 | 32.57/1.14 | 30.79/1.51 | 32.91/0.92 |
| Helmet | 25.08/1.22 | 26.95/1.09 | 28.78/1.31 | 32.62/0.23 | 30.06/0.15 |
| Toaster | 18.66/0.72 | 23.95/0.75 | 20.07/0.80 | 25.89/0.53 | 25.02/0.68 |
| Teapot | 38.21/—— | 43.35/—— | 43.86/—— | 43.88/—— | 43.39/—— |

Table 7. Comparison in terms of PSNR and CD↓ ($\times 100$).

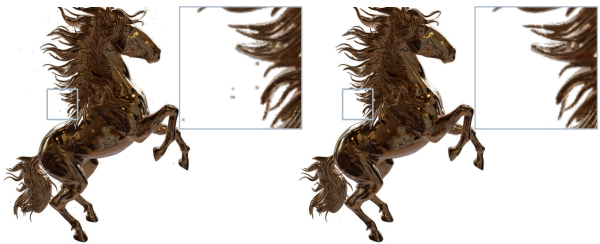

Figure 1. Without the projection-based consistency loss, we observe the artifacts caused by outliers (left). The loss encourages the outlier towards the surface, thus diminishing the artifacts.

### 3.1. Extra ablation studies

We ablate our hyperparameter choice and loss usage in this part, including the threshold in the projection-based loss, the impact of the median and the mask loss.

**The choice of the threshold $\varepsilon$.** A large threshold forces Gaussians onto wrong surfaces, while a small one causes most Gaussians to be excluded, resulting in decreased performance. Therefore, we choose to set $\varepsilon = 0.1$ for all scenes, which reveals its generalization ability and yields the best performance in Tab. 5.

**The impact of the median loss.** We use the median loss to encourage $\gamma$ towards $\gamma_\mathrm{m}$, instead of directly setting $\gamma = \gamma_\mathrm{m}$, because $\gamma_\mathrm{m}$ is used to give the direction of optimization and not the target. Besides, the median loss enables the further optimization of the transformation. Our experiment also proves that the median loss promotes the convergence while others variants fail, as shown in Tab. 6. Besides, the setting of $\mathcal{T}_{\gamma_\mathrm{m}}(|s|_\mathrm{m}) = o_m = 0.5$ is from our observation, as larger $\gamma_m$ values cause early over-pruning of Gaussians, while smaller values fail to narrow the transformation. Both cases reduce performance.

**The impact of the mask loss.** While removing the mask loss leads to a performance decrease of 1dB, it still outperforms GS-ROR trained with the mask loss, as shown in Tab. 4.

### 3.2. Extra benefits from the projection-based loss

Our projection loss also promotes the Gaussian outliers moving toward the surface and diminishes the artifacts, as shown in Fig. 1.

### 3.3. Glossy Blender dataset

We present the relighting results with the decomposition of material and geometry on the Glossy Blender dataset [6] in Fig. 2. As there is no material ground truth, we only provide the visualization of our method. More qualitative comparisons are in Fig. 3. Our method offers smooth geometry with details preserved and thus outperforms other Gaussian-based methods regarding relighting quality. Besides, we compare our method with four NeRF-based inverse rendering methods in Tab. 2, and our method is still competitive. MII and TensoIR do not design for reflective surfaces and thus perform poorly on the Glossy dataset. Although our method has a performance gap compared to NeRO and TensoSDF, our method is far more efficient than theirs with 50% memory usage and 17% training time at most. Our method also supports real-time rendering while other NeRF-based methods fail[1]. Besides, as shown in Fig. 4, the results from our methods include sharp details that other NeRF-based methods fail to reconstruct and thus are more visually realistic.

### 3.4. Shiny Blender dataset

We present the decomposition of material and geometry with the relighting results on the Shiny Blender dataset [8] in Fig. 5. Due to no relighting ground truth, we show the environment map for shading as the reference. Besides, we present more qualitative comparisons of normal in Fig. 6. Due to the robustness of our method, our method provides high-quality relighting results for reflective surfaces. Additionally, We present the evaluation of surface quality in terms of Chamber distance (CD) along with NVS results in Tab. 7. Our method outperforms other methods in terms of CD, while GS-ROR achieves a higher quality than ours in some cases. The main reason is that our backbone is 2DGS

---
[1] NeRO and TensoSDF use the Cycles Render Engine in Blender to provide relighting results, so the FPS depends on samples per pixel, which is 1024 following the NeRO setting.

while theirs is 3DGS. Although 2DGS demonstrates superior reconstruction quality, its NVS quality is lower compared to that of 3DGS.

### 3.5. TensoIR synthetic dataset

We present the decomposition of material and geometry on the TensoIR synthetic dataset [2] in Fig. 7. Besides, we present more qualitative comparisons of normal in Fig. 8. Our method provides a reasonable decomposition for diffuse objects and realistic relighting results. The quantitative comparison between our method and NeRF-based methods is shown in Tab. 3, and the qualitative comparison between our method and NeRF-based methods is Fig. 9. Our method is numerically and visually competitive for diffuse object relighting while maintaining efficiency.

### 3.6. NeILF++ dataset

We present more comparisons between our method and other Gaussian-based inverse rendering methods on the NeILF++ [9] dataset, revealing the robustness of our method for real-world objects. As shown in Fig. 10, our method provides realistic relighting results and detailed normal for diverse real objects.

## References

[1] Yingwenqi Jiang, Jiadong Tu, Yuan Liu, Xifeng Gao, Xiaoxiao Long, Wenping Wang, and Yuexin Ma. GaussianShader: 3D Gaussian splatting with shading functions for reflective surfaces. In *CVPR*, pages 5322–5332, New York, NY, USA, 2024. IEEE. 1

[2] Haian Jin, Isabella Liu, Peijia Xu, Xiaoshuai Zhang, Songfang Han, Sai Bi, Xiaowei Zhou, Zexiang Xu, and Hao Su. TensoIR: Tensorial inverse rendering. In *CVPR*, pages 165–174, New York, NY, USA, 2023. IEEE. 2, 4

[3] Brian Karis. Real shading in unreal engine 4. *Proc. Physically Based Shading Theory Practice*, 2013. 1

[4] Bernhard Kerbl, Georgios Kopanas, Thomas Leimkühler, and George Drettakis. 3D Gaussian splatting for real-time radiance field rendering. *ACM TOG*, 42(4):139–1, 2023. 1

[5] Jia Li, Lu Wang, Lei Zhang, and Beibei Wang. TensoSDF: Roughness-aware tensorial representation for robust geometry and material reconstruction. *ACM TOG*, 43(4), 2024. 2

[6] Yuan Liu, Peng Wang, Cheng Lin, Xiaoxiao Long, Jiepeng Wang, Lingjie Liu, Taku Komura, and Wenping Wang. NeRO: Neural geometry and BRDF reconstruction of reflective objects from multiview images. *ACM TOG*, 42(4), 2023. 2, 3

[7] Stephen McAuley, Stephen Hill, Adam Martinez, Ryusuke Villemin, Matt Pettineo, Dimitar Lazarov, David Neubelt, Brian Karis, Christophe Hery, Naty Hoffman, and Hakan Zap Andersson. Physically based shading in theory and practice. In *ACM SIGGRAPH 2013 Courses*, New York, NY, USA, 2013. Association for Computing Machinery. 1

[8] Dor Verbin, Peter Hedman, Ben Mildenhall, Todd Zickler, Jonathan T. Barron, and Pratul P. Srinivasan. Ref-NeRF: Structured view-dependent appearance for neural radiance fields. In *CVPR*, pages 5491–5500, New York, NY, USA, 2022. IEEE. 3

[9] Jingyang Zhang, Yao Yao, Shiwei Li, Jingbo Liu, Tian Fang, David McKinnon, Yanghai Tsin, and Long Quan. NeILF++: Inter-reflectable light fields for geometry and material estimation. In *ICCV*, pages 3601–3610, New York, NY, USA, 2023. IEEE. 4

[10] Yuanqing Zhang, Jiaming Sun, Xingyi He, Huan Fu, Rongfei Jia, and Xiaowei Zhou. Modeling indirect illumination for inverse rendering. In *CVPR*, pages 18643–18652, New York, NY, USA, 2022. IEEE. 2

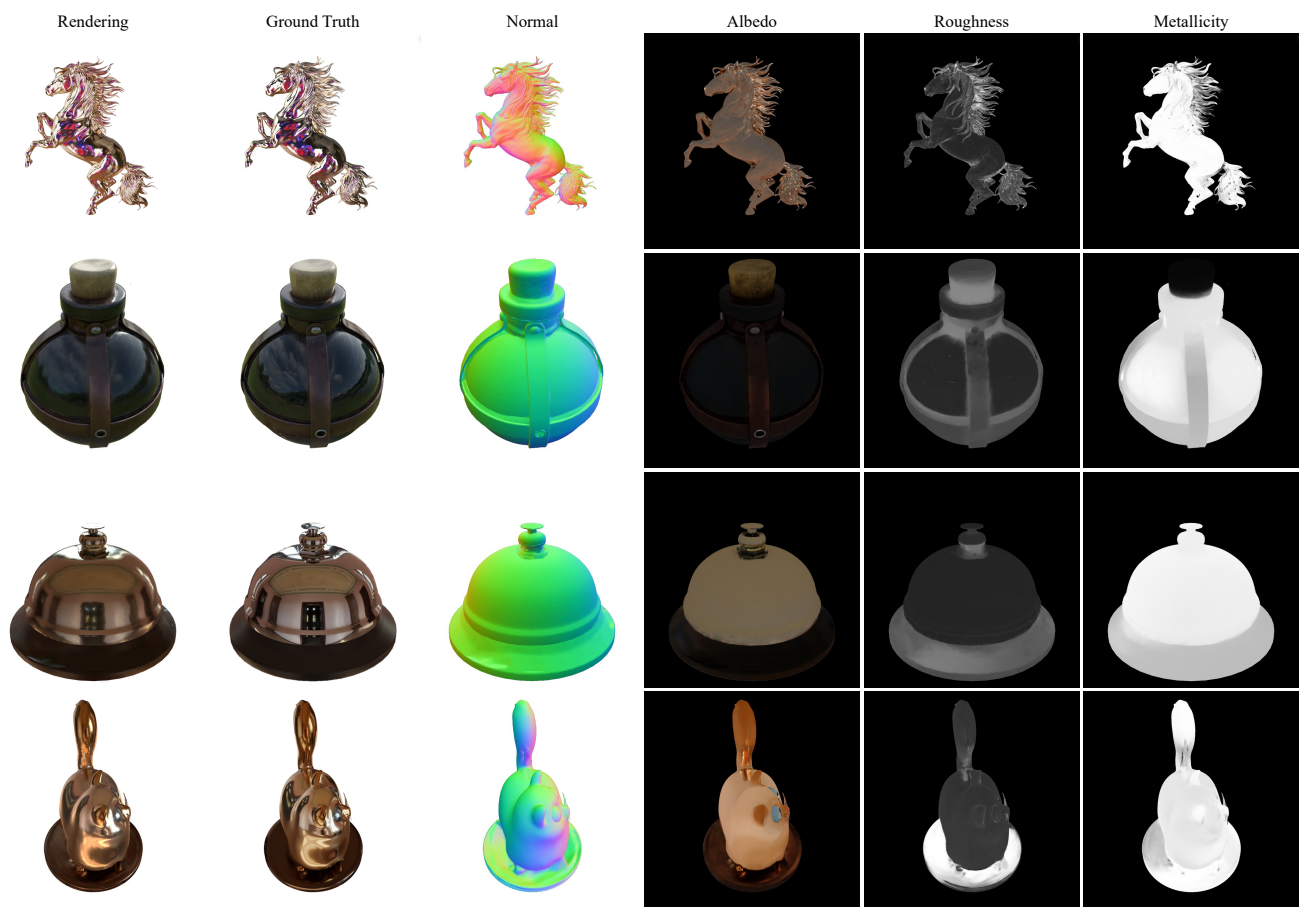

Figure 2. Decomposed maps of our method on the Glossy Blender dataset. Our method can provide a reasonable decomposition for reflective surfaces.

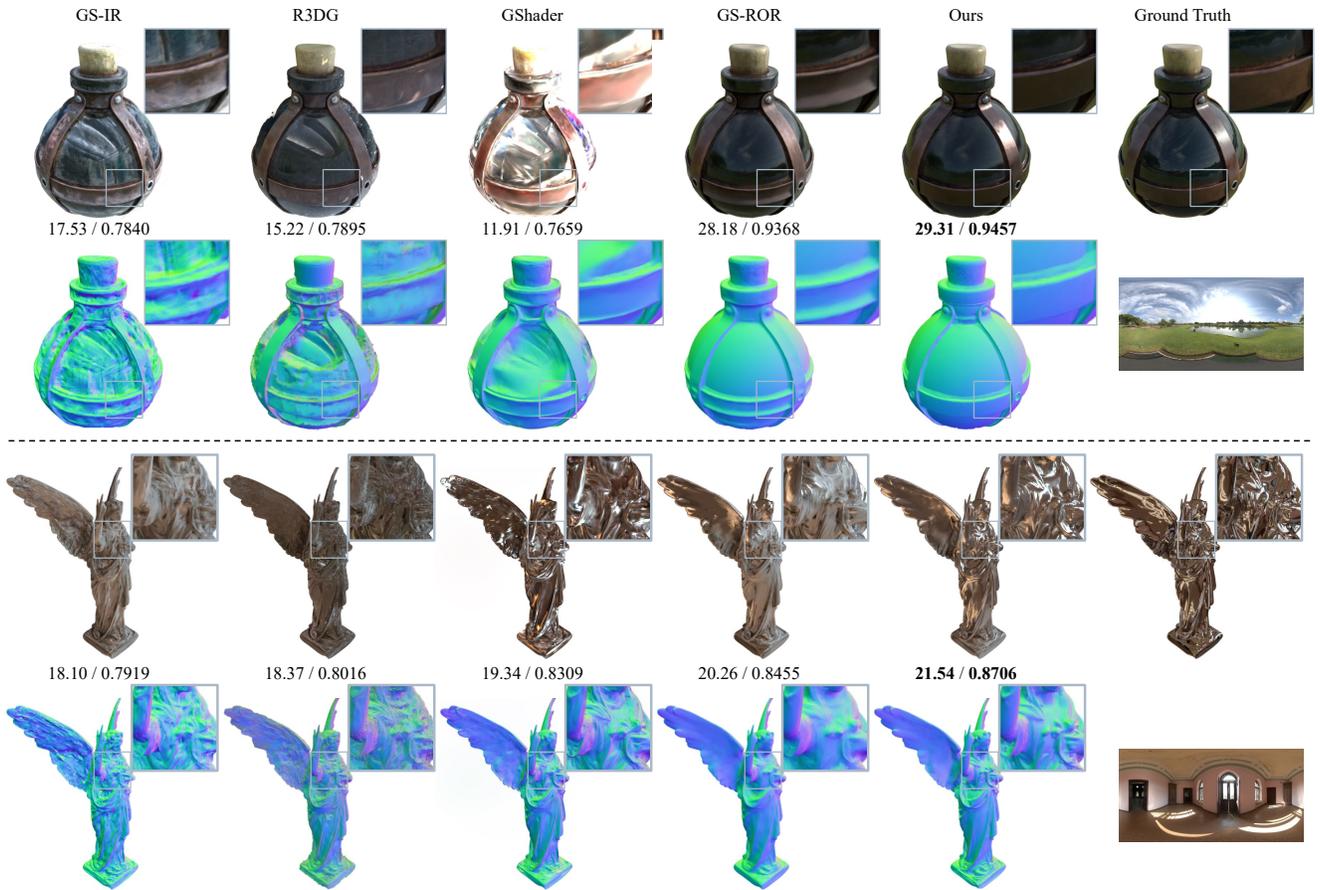

Figure 3. The qualitative comparison with Gaussian-based methods in terms of relighting results and normal on the Glossy Blender dataset. Our method provides high-quality relighting results for reflective surfaces. The PSNR/SSIM of relighting results under the current view are below the images.

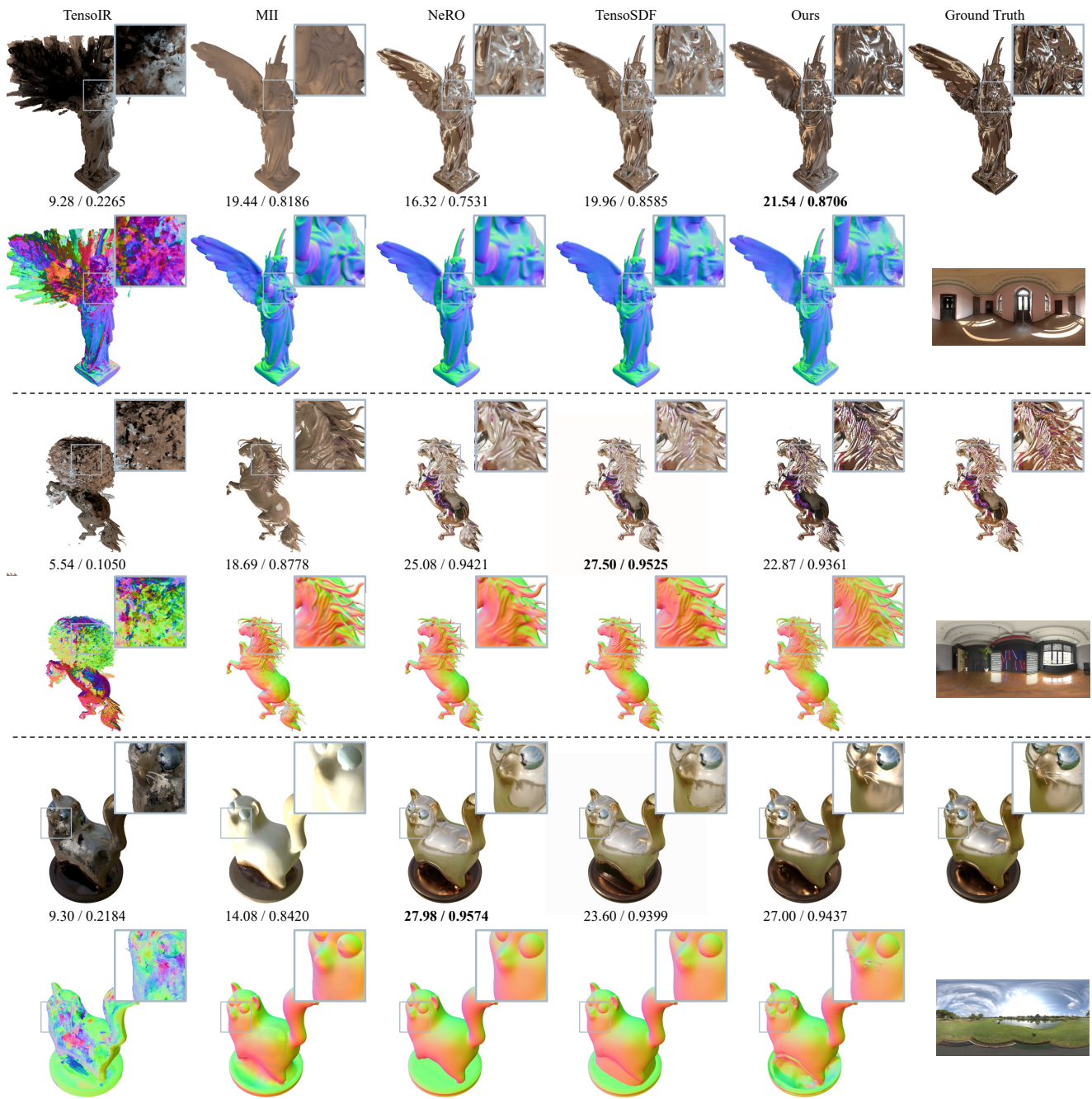

Figure 4. The qualitative comparison with NeRF-based methods in terms of relighting results and normal on the Glossy Blender dataset. Our method is competitive with these methods and outperforms in the detailed regions. The PSNR/SSIM of relighting results under the current view are below the images.

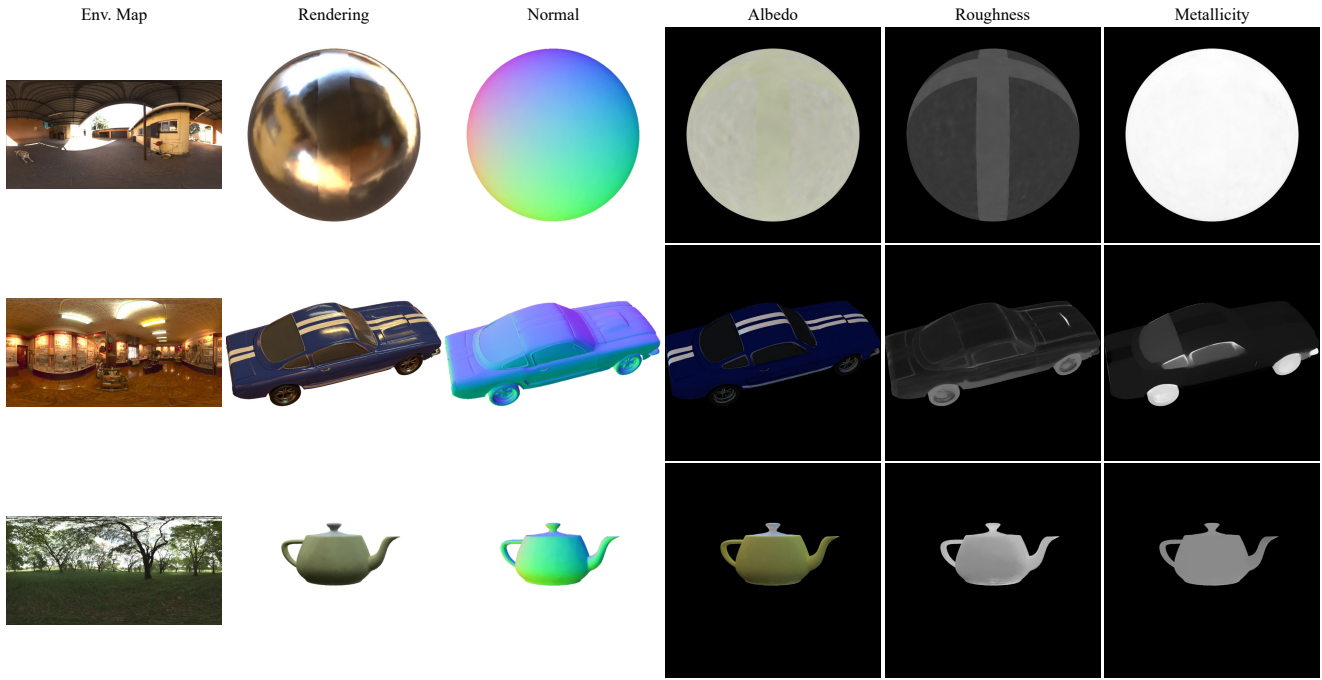

Figure 5. Decomposed maps of our method on the Shiny Blender dataset. The relighting ground truth is unavailable on the dataset, so we provide the environment map for shading as a reference.

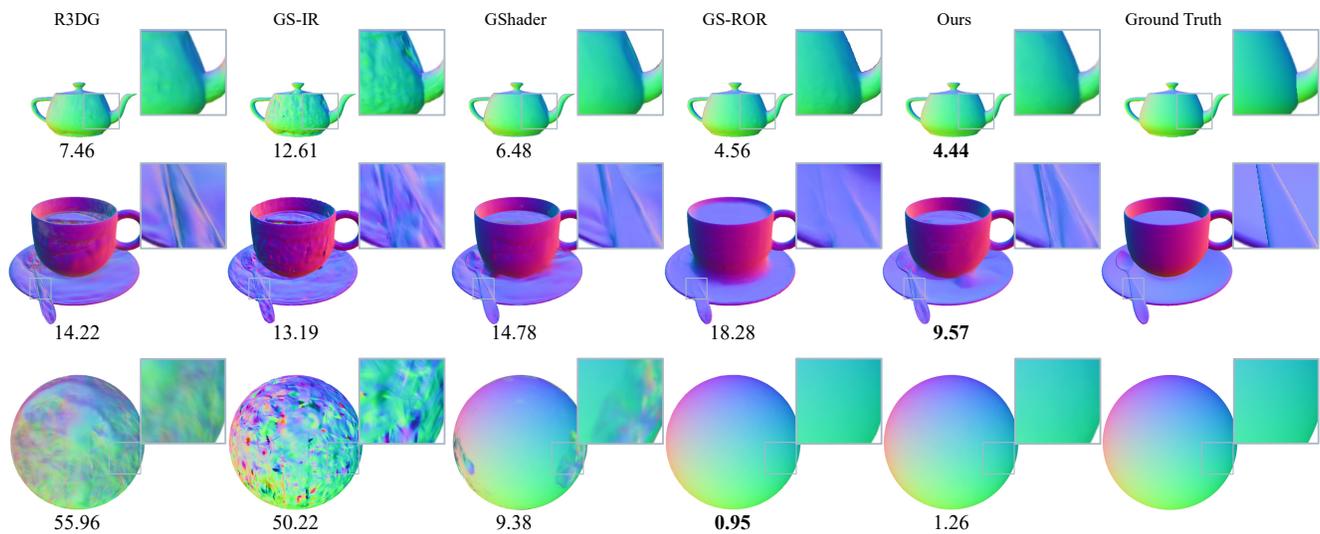

Figure 6. The qualitative comparison with Gaussian-based methods in terms of normal on the Shiny Blender dataset. The MAE under the current view is above the visualization. Our method can provide reasonable normal for diverse objects.

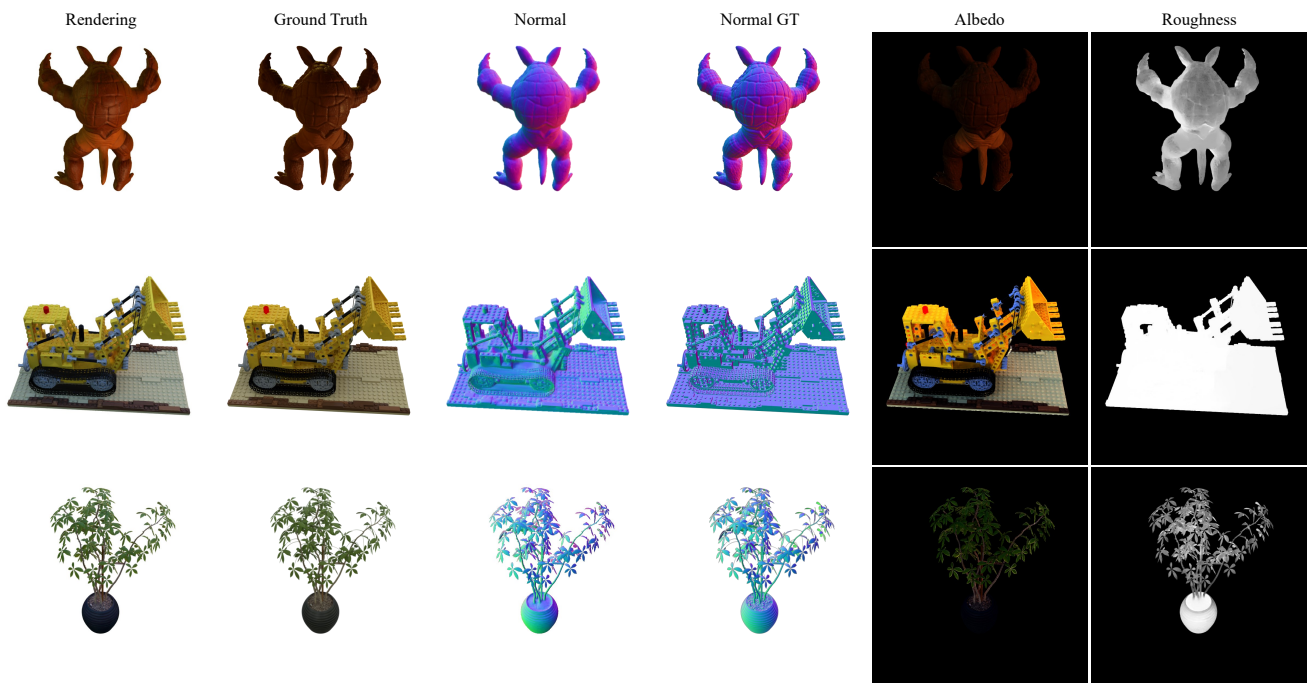

Figure 7. Decomposed maps of our method on the TensoIR synthetic dataset. Our method can provide a reasonable decomposition for diffuse objects.

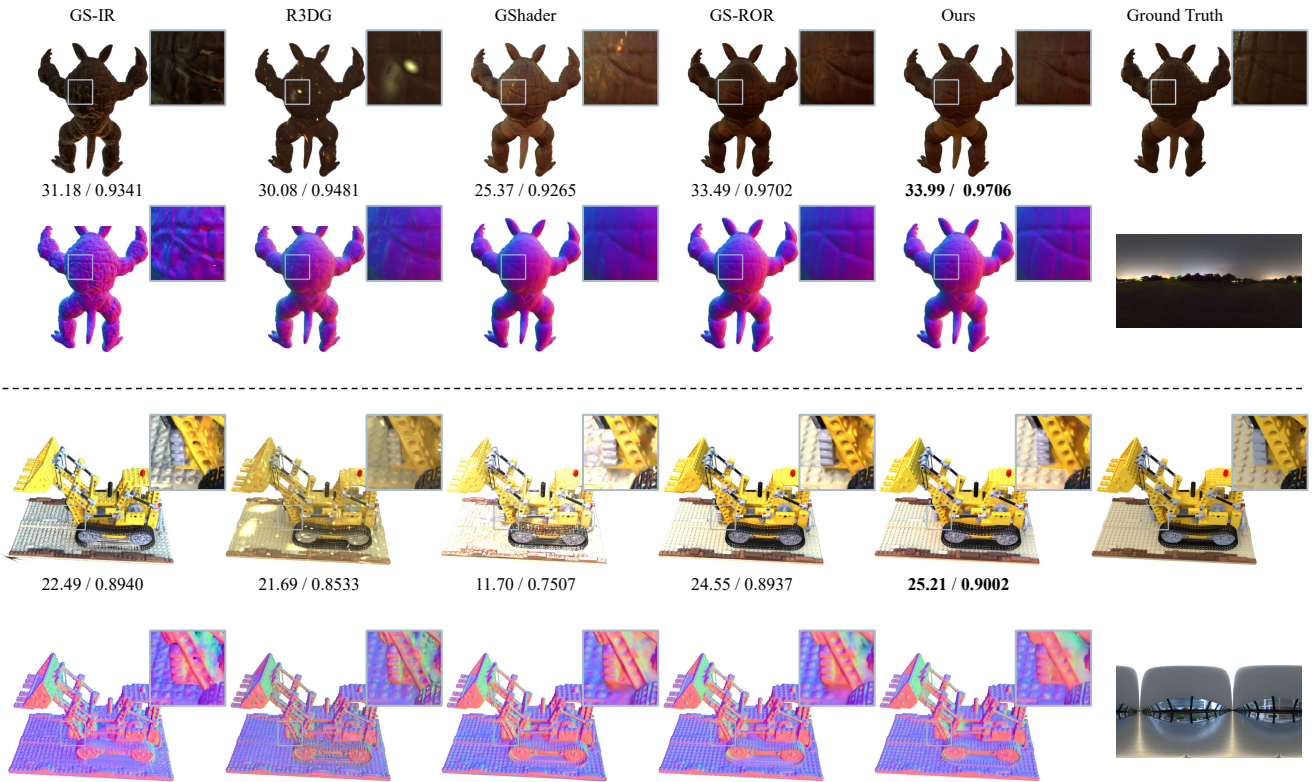

Figure 8. The qualitative comparison with Gaussian-based methods in terms of relighting results and normal on the TensoIR synthetic dataset. Our method can provide robust normal for diffuse objects. The PSNR/SSIM of relighting results under the current view are below the images.

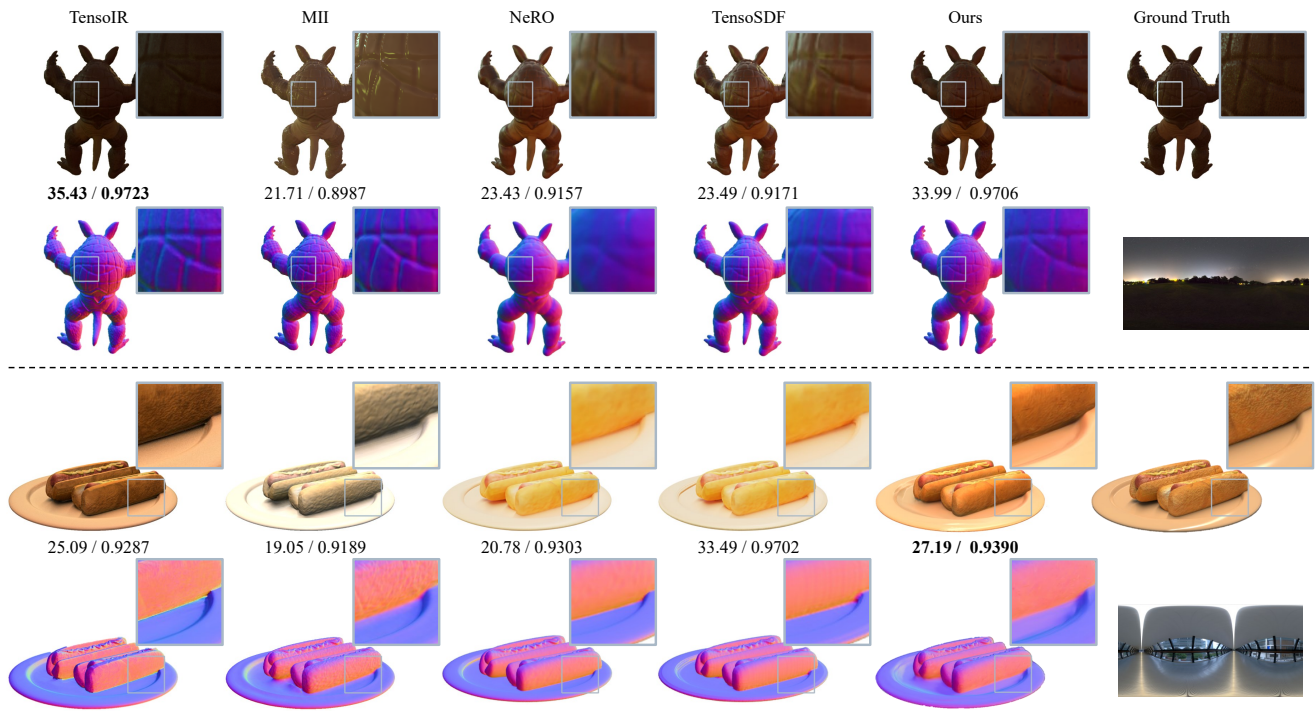

Figure 9. The qualitative comparison with NeRF-based methods in terms of relighting results and normal on the TensoIR synthetic dataset. Our method is competitive while maintaining high efficiency. The PSNR/SSIM of relighting results under the current view are below the images.

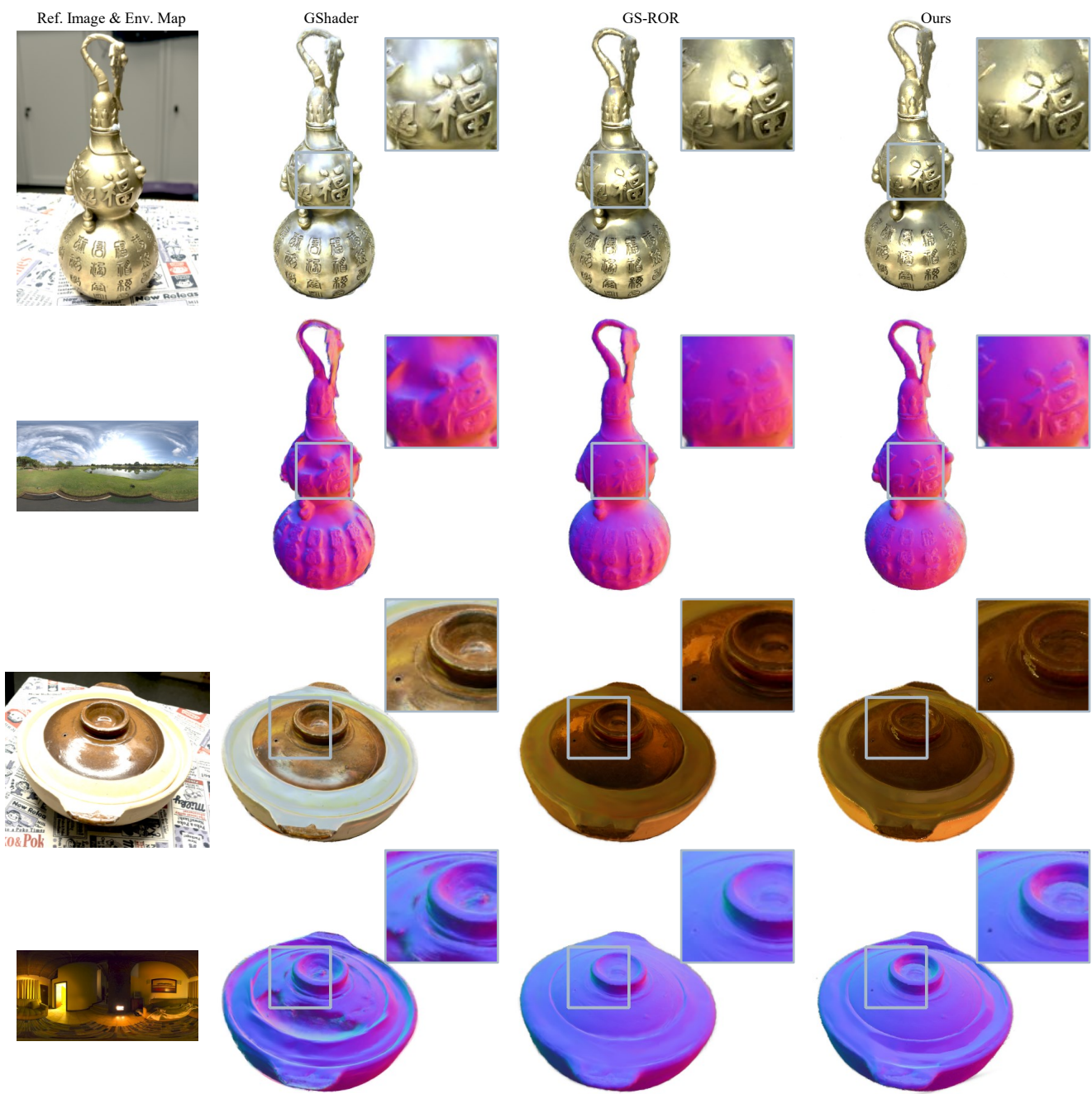

Figure 10. The qualitative comparison in terms of relighting results and normal on the NeILF++ dataset. Our method ensures the smoothness of normal while preserving details for real data, thus providing more realistic relighting results.


\end{document}